\documentclass[lettersize,journal]{IEEEtran}
\usepackage{amsmath,amsfonts}
\usepackage{algorithmic}
\usepackage{algorithm}
\usepackage{array}
\usepackage[caption=false,font=normalsize,labelfont=sf,textfont=sf]{subfig}
\usepackage{textcomp}
\usepackage{stfloats}
\usepackage{url}
\usepackage{verbatim}
\usepackage{graphicx}
\graphicspath{{./figures/}}

\usepackage[sort&compress,comma,numbers]{natbib}
\usepackage{hyperref}
\usepackage{color,soul}
\usepackage{bm}
\usepackage{multirow}
\usepackage{booktabs}

\usepackage{tikz}
\usetikzlibrary{shapes.geometric}
\usetikzlibrary{decorations.pathreplacing}

\begin{document}

\newcommand{\tcb}{\textcolor{black}}

\title{Survey on Visual Signal Coding and Processing with Generative Models: Technologies, Standards and Optimization}

\author{Zhibo Chen, Heming Sun, Li Zhang and Fan Zhang
\thanks{Zhibo Chen is with University of Science and Technology of China, China (e-mail: chenzhibo@ustc.edu.cn), Heming Sun is with Yokohama National University, Japan (email: sun-heming-vg@ynu.ac.jp), Li Zhang is with Bytedance Inc., US (e-mail: lizhang.idm@bytedance.com), Fan Zhang is with University of Bristol, UK (email: fan.zhang@bristol.ac.uk).}}

\maketitle

\begin{abstract}
This paper provides a survey of the latest developments in visual signal coding and processing with generative models. Specifically, our focus is on presenting the advancement of generative models and their influence on research in the domain of visual signal coding and processing. This survey study begins with a brief introduction of well-established generative models, including the Variational Autoencoder (VAE) models, Generative Adversarial Network (GAN) models, Autoregressive (AR) models, Normalizing Flows and Diffusion models. The subsequent section of the paper explores the advancements in visual signal coding based on generative models, as well as the ongoing international standardization activities. In the realm of visual signal processing, our focus lies on the application and development of various generative models in the research of visual signal restoration. We also present the latest developments in generative visual signal synthesis and editing, along with visual signal quality assessment using generative models and quality assessment for generative models. The practical implementation of these studies is closely linked to the investigation of fast optimization. This paper additionally presents the latest advancements in fast optimization on visual signal coding and processing with generative models. We hope to advance this field by providing researchers and practitioners a comprehensive literature review on the topic of visual signal coding and processing with generative models.
\end{abstract}


\section{Introduction}
In recent years, generative models have emerged as one of the most significant and rapidly developing areas of research in artificial intelligence. Generative models have demonstrated remarkable success in synthesizing high-quality data (text, image, video, 3D content, etc.) and hold promise for utilizing unlabeled data, transfer learning, data augmentation, drug/protein generation, and other applications. Different generative models have been developed to capture complex data distributions and generate new examples. For example, the Autoregressive (AR) models, like GPT-4, sequentially predict and generate data. The Variational Autoencoders (VAE) learn the parameters of a probability distribution representing the input data. The Generative Adversarial Networks (GAN) train two competing neural networks in an adversarial manner to generate realistic synthetic data. The flow models learn invertible mappings between data and latent space. The Diffusion models iteratively add noise to data and then train a neural network to reverse the process. 

Simultaneously, generative models have been demonstrated to be a crucial tool for learning-based visual signal coding and processing. For example, the VAE model has been widely employed as a foundational framework in end-to-end learning-based image coding schemes. The AR model has been extensively studied to improve entropy coding performance, and the GAN model and Diffusion model have been utilized frequently to enhance the subjective quality of coding schemes. Additionally, generative models have also been explored in various visual signal processing tasks, including restoration, enhancement, editing, quality assessment, and interpolation.

In light of the rapid growth of visual signal coding and processing with generative models, its contributions to international standards and practical application optimization are increasingly valued. The Joint Video Experts Team (JVET) of \textcolor{black}{International Telecommunication Union - Telecommunication Sector Video Coding Experts Group (ITU-T VCEG) and International Organization for Standardization\slash International Electrotechnical Commission} Moving Picture Experts Group (ISO/IEC MPEG) has started working together on an exploration study to evaluate potential neural network-based video coding (NNVC) technology beyond the capabilities of the conventional hybrid video coding framework as early as 2018. In addition, MPEG also launched many standardization projects, which have started adopting artificial intelligence (AI)-based technologies, such as AI-based 3D graphics coding, AI model compression, and video coding for machines (VCM). Meanwhile, the Joint Photographic Experts Group (JPEG) AI, the creation of a learning-based image coding standard and a joint standardization project {\color{black}between ISO/IEC JPEG and ITU-T}, is being developed. It is observed that these standards built on AI-based methods are able to serve various purposes, including significant compression efficiency improvement, efficient multimedia representation, compression and deployment of neural network models, metadata extraction for AI-based processes, and multimedia content processing. Although learning-based visual signal coding has obtained a remarkable coding gain, the complexity is still too high to achieve real-time processing. Regarding the algorithm, the AR model is usually used to fully utilize the correlation of neighboring samples. In addition, the huge scale of the neural network increases the burden of computational and memory cost. Regarding hardware and implementation, most frameworks are accelerated by general-purpose accelerators such as Nvidia GPU and AMD/Xilinx DPU. However, the hardware utilization of GPU/DPU is relatively low since the computation-to-communication (CTC) ratio of neural network\tcb{s} may not fit the hardware resources such as the number of PE core\tcb{s} and bandwidth. Recently, transformer-like architectures \tcb{have} gradually become popular due to their superior performance. To accommodate the rapid \tcb{development} of neural network architectures, developing a generic and efficient hardware accelerator is also a challenge. 

Therefore, we seek to further advance this important research area by providing researchers and practitioners with a broad reference to the literature on visual signal coding and processing with generative models. The remainder of this paper provides a brief overview of the visual signal coding and processing with generative models from the perspectives of the background of generative models, the development of visual signal coding with generative models and related {\color{black}standardization activities}, and the recent progress of visual signal processing with generative models. We start with an overview of generative models in Section \ref{sec:GM} and then introduce visual signal coding with generative models in Section \ref{sec:coding}, with an introduction of {\color{black}standardization activities on visual signal coding with generative models} in Section \ref{sec:Stan}. \textcolor{black}{In Section \ref{sec:GVSP}, we review some ongoing research activities related to visual signal processing with generative models, including restoration, enhancement, editing, and interpolation. The quality assessment methods based on generative models and those for generative models are further discussed in Section \ref{sec:QA}}. Finally, a review of fast implementation and optimization of visual signal coding and processing with generative models is presented in Section \ref{sec:Optimization}, followed by a conclusion in Section \ref{sec:Conclu}.

\section{Generative Models}
\label{sec:GM}

This section provides a brief introduction to generative models, including GANs, VAEs, Autoregressive Models, Normalizing Flows, and Diffusion models. 
\subsection{Generative Adversarial Networks (GAN)}

\tcb{
GANs~\cite{goodfellow2014generative} are important components of deep generative models. They are developed to generate data through an adversarial training strategy involving both generators and discriminators. In this context, the objective of the generator is to generate data samples that are as realistic as possible to deceive the discriminator. The discriminator is tasked with differentiating these generated data samples from the actual ones in the training data, and both networks are updated iteratively. The objective is to find an equilibrium point where the discriminator cannot reliably discern fake data samples from real data samples. Utilizing the discriminator's ability to understand perception, GANs achieve outstanding results in qualitative generation and are widely used in a wide range of tasks, including conditional generation~\cite{mirza2014conditional}, representation learning~\cite{chen2016infogan}, image-to-image translation~\cite{I2ISurvey}, image super-resolution~\cite{ledig2017photo}, image enhancement~\cite{chen2018deep}, style transfer~\cite{karras2019style} and semantic editing~\cite{shen2020interfacegan}. GANs are also capable of generalizing to generate data in modalities other than images, such as video~\cite{tulyakov2018mocogan}, text~\cite{yu2017seqgan}, audio~\cite{kong2020hifi} and 3D data~\cite{SPGAN}. However, adversarial training can be difficult due to issues such as instability and mode collapse \cite{arjovsky2017towards}, which lead to low-quality generation outputs with limited variability. Some research \cite{arjovsky2017wasserstein} has attempted to mitigate this issue by employing more rational loss functions.}

\subsection{Variational Autoencoders (VAE)}
\tcb{
VAEs~\cite{kingma2013auto} are a type of generative model that employ Bayesian inference to approximate the distribution of data. VAEs comprise two components: an encoder that transforms input data into a distribution of latent variables, and a decoder that reconstructs input data from the latent variables. The training objective of VAEs is to optimize a lower bound on the data log-likelihood, which is composed of two terms: a Kullback–Leibler (KL) divergence term that quantifies the dissimilarity between the distribution of latent variables and the prior distribution, and a reconstruction term that measures the fidelity of the generated data. VAEs have been improved by addressing several challenges, such as increasing the expressiveness of the latent variable distribution~\cite{DBLP:journals/corr/BurdaGS15}, reducing the gap between the lower bound and the true log-likelihood, avoiding posterior collapse~\cite{DBLP:conf/iclr/RazaviOPV19}, and scaling up to high-resolution data \cite{vahdat2020nvae}. Some of the notable works that have contributed to these improvements are IAF-VAE~\cite{Kingma2016ImprovingVA}, NVAE~\cite{vahdat2020nvae}, and VDVAE~\cite{child2020very}, which have enhanced the performance and quality of VAEs, making them more powerful for generative modeling. Meanwhile, VAEs are increasingly used as part of other generative models, such as normalizing flows and diffusion models. These combinations have the added benefit of enhancing the performance of the generated data samples.}

\subsection{Autoregressive Models}
\tcb{Autoregressive models~\cite{ostrovski2018PixelIQN} view generation as a sequential process, predicting future outcomes based on previously observed data. They excel in precision, optimizing the likelihood of the estimated data by learning dependencies within the sequence. This is typically achieved through a masking strategy, such as PixelRNN~\cite{van2016pixel} and Gated PixelCNN~\cite{van2016conditional}, where certain known values are used to predict unknown neighboring values. Autoregressive models exhibit exceptional performance in modeling distribution density and capturing intricate patterns in data. However, their sequential nature results in a slow data generation process~\cite{bond2021DeepGenerativeModel}. Furthermore, by relying on historical data to make predictions about the future, they run the risk of overfitting the training set and potentially generating duplicates of the observations. Recent studies~\cite{hoogeboom2021autoregressive, wu2024ArDiffusion} have demonstrated that integrating autoregressive models with diffusion models~\cite{ho2020denoising} substantially improves their generation speed, thereby enhancing the performance of autoregressive models.
}

\subsection{Normalizing Flows}
\tcb{Normalizing flows \cite{rezende2015variational,kobyzev2020normalizing} are a class of generative models designed to transform complex data distributions into simpler, more tractable forms such as Gaussian distributions. This is achieved through a series of invertible transformation layers, and by stacking such layers, normalizing flows are able to map an intricate distribution into a simpler one~\cite{grathwohl2018ffjord}. The prerequisite for the invertibility of a transformation layer is crucial, as it fulfills two functions: it should permit the transformation of complex data into a more manageable distribution for analysis purposes, and it must also facilitate the creation of new data instances from this simplified distribution. To optimize the model, a tractable marginal likelihood is computed, which requires each transformation layer to be capable of calculating its Jacobian determinant efficiently. Some notable works include RealNVP~\cite{dinh2016density}, GLOW~\cite{kingma2018glow}, and Residual Flow~\cite{chen2019ResidualFlow}. Normalizing flows are noted for their capacity to learn features and quick generation process. However, the stringent requirement for transformation modules to be invertible often makes it hard to choose more flexible network structures, resulting in inferior quantitative performance in density modeling. Despite this, their ability to perform exact likelihood calculations makes them a valuable tool for various tasks, including sample generation, latent variable projection, and density value estimation~\cite{albergo2022building, zhang2021DiffFlow,zand2023DiNof}. }

\subsection{Diffusion Models}
\tcb{Diffusion models have achieved noteworthy accomplishments within the field of generative models. It first attracted extensive interest and widespread recognition with the publication of the paper titled `Denoising Diffusion Probabilistic Models' ~\cite{ho2020denoising} in 2020. Similar ideas also came to the attention of the public when Score-based Generative Models~\cite{song2020score} were proposed, bridging both the diffusion model and score-based generative model into a unified framework. There are endeavors focusing on theoretical or engineering optimization such as accelerating sampling speed, like DDIM~\cite{song2020denoising} and DPM-Solver~\cite{lu2022dpm}, or cutting down training cost, like Stable Diffusion~\cite{rombach2022high}. These  works enhance the practical performance of diffusion models, making them more tractable for either training or inference. Concurrent with its rapid development, the diffusion model has emerged as a prominent generative model known for its strong theoretical foundation and exceptional performance. It has been widely applied in various downstream applications like image inpainting~\cite{lugmayr2022repaint}, image-to-image translation~\cite{parmar2023zero}, image composition~\cite{lu2023tf}, image customization~\cite{gal2022image} and prompt editing~\cite{gal2022image}. ControlNet ~\cite{zhang2023adding} was proposed to additionally integrate various applications within a single framework. Beyond images, diffusion models also succeed in generating contents of other modalities, including videos~\cite{blattmann2023stable} and 3D objects~\cite{poole2022dreamfusion}.}

\section{Visual Signal Coding with Generative Models}
\label{sec:coding}

This section provides a concise review of the application of generative models to visual signal coding, focusing primarily on image and video coding. 

\subsection{Image Coding with Generative Models}
In fact, the phrase ``image coding with generative models" can have multiple meanings. On the one hand, probabilistic generative models provide the theoretical foundations for end-to-end learned image coding, sometimes referred to as learning-based image coding, neural network-based image coding, or neural image coding in the literature. Specifically, probabilistic generative models, such as variational autoencoders (VAEs) \cite{kingma2013auto} and diffusion models \cite{sohl2015deep} contribute to successful frameworks for neural image coding \cite{balle2016end,theis2022lossy}. Also, probabilistic generative models, such as autoregressive models \cite{van2016conditional} and normalizing flows \cite{rezende2015variational}, inspire several important improvements in coding performance \cite{minnen2018joint,xie2021enhanced}. On the other hand, some well-established generative models, such as Generative Adversarial Networks (GANs) \cite{goodfellow2014generative} and diffusion models, can be combined with these end-to-end learned image coding models, which have been demonstrated to provide better perceptual quality. In this section, we overview techniques in both areas, and review an important theory in the field of generative image coding: the rate-distortion-perception tradeoff.

\subsubsection{Probabilistic Generative Models for Image Coding} \label{sec:framework_genertive_model}
Prevalent methods for neural image compression follow a variational autoencoder framework. 
\textcolor{black}{Specifically, the input image $\bm{x}$ is usually mapped into its latent representations $\bm{y}$, which are then quantized into $\bm{\hat{y}}$.
 Since the gradient of the scalar quantization function is zero almost everywhere \cite{guo2021soft}, most methods for neural image compression employ additive uniform noise to approximate quantization during training. Early works \cite{balle2016end,balle2018variational} connect the rate-distortion objective and variational inference in this noise-relaxed case, 
\begin{equation}
\begin{aligned}
    \mathbb{E}_{\bm{x}\sim p_{\bm{x}}} & D_{\mathrm{KL}}  (q(\bm{\tilde{y}}|\bm{x})|p(\bm{\tilde{y}}|\bm{x}))  =  \mathbb{E}_{\bm{x}\sim p_{\bm{x}}} \log p(\bm{x}) \  + \\
     \mathbb{E}_{\bm{x}\sim p_{\bm{x}}} & \mathbb{E}_{\bm{\tilde{y}}\sim q_{\bm{\tilde{y}}|\bm{x}}} [\log q(\bm{\tilde{y}}|\bm{x}) -   \log  p(\bm{x} |\bm{\tilde{y}})-\log p (\bm{\tilde{y}})].
\end{aligned}
\label{equation_variational_theory}
\end{equation}
Since the image $\bm{x}$ is given in the task of compression, the first right-hand-side term in the above equation is a constant during optimization. The second right-hand-side term evaluates to zero as we employ additive standard uniform noise as a stand-in for quantization during training: 
\begin{equation}
q(\boldsymbol{\tilde{y}}|\boldsymbol{x})=q(\boldsymbol{\tilde{y}}|\boldsymbol{y})= \mathcal{U}(\boldsymbol{\tilde{y}}|\boldsymbol{y}-0.5, \boldsymbol{y}+ 0.5)=1.
\label{equation2}
\end{equation}
The rest two terms in Eq. \eqref{equation_variational_theory} denote the distortion and rate, respectively. Therefore, the rate-distortion optimization objective in lossy compression can be successfully interpreted from the view of the variational inference.
Such a joint rate-distortion optimization objective is critical to achieve promising compression performance. 
Despite a short history, the rate-distortion performance of this variational image compression framework has been demonstrated to surpass traditional image compression standards, in terms of both the objective metrics such as RGB PSNR \cite{guo2021causal} or MS-SSIM \cite{he2022elic}, and perceptual quality \cite{mentzer2020high}. In addition to noise-relaxed quantization, some alternatives, e.g. vector quantization \cite{agustsson2017soft,feng2023nvtc}, and soft-to-hard annealing \cite{yang2020improving}, have been proposed to replace additive uniform noise during training. However, the mainstream approaches for neural image compression still adopt additive uniform noise during training, since it is stable during training and theoretically sound as variational autoencoders.  Soft-then-hard two stage quantization \cite{guo2021soft} scheme was further proposed to learn an expressive latent space softly, then closes the train-test mismatch with hard quantization.} 

Recently, as a new member of probabilistic generative models, diffusion models have attracted increasing attention due to their strong capability to model data distributions \cite{kingma2021variational}. Lucas et al. \cite{theis2022lossy} proposed a promising framework that applies Denoising Diffusion Probabilistic Models (DDPMs) \cite{ho2020denoising} to neural image coding, where the image information is decomposed into posterior distributions in multiple diffusion steps and compressed with relative entropy coding \cite{flamich2020compressing,flamich2022fast}. This new framework exhibits encouraging compression performance in terms of both objective metrics and perceptual quality. Diffusion models operate by iteratively transforming an input image into a noise distribution, allowing for effective compression while preserving essential features. However, due to the inherent huge complexity of DDPMs and relative entropy coding, this diffusion-based lossy compression framework still has ample room for further improvement. 

Autoregressive models for image generation were first proposed as PixelCNN \cite{van2016conditional}. \textcolor{black}{Although generation and compression are fundamentally two different tasks, the concept of autoregression inspires the design of the context model, which significantly enhances the compression performance of of neural image compression models.} 
By decoding the latent variables sequentially in raster-scan order, the context model can improve the entropy modeling of latent variables, and therefore boost the compression performance. Subsequently, the spatial autoregressive context model was proposed to be replaced by the channel autoregressive context model \cite{minnen2020channel}, which leverages the correlation among latent channels and is more efficient for decoding. 

Another typical type of probabilistic generative models is normalizing flows. Although the applications of normalizing flows in neural image compression are not as many as the previous three categories of probabilistic generative models, normalizing flows still play an important role for image compression. For example, normalizing flows can help the neural image compression model to achieve better idempotence \cite{li2023idempotent}, which means that a codec can support successive image compression \cite{helminger2020lossy}. The ANFIC \cite{ho2021anfic} and its subsequent studies \cite{ho2022canf, chen2023b, alexandre2023hierarchical} offered a comprehensive analysis of applying Normalizing Flows to lossy image and video compression. The iWave work \cite{ma2019iwave} proposed a wavelet-like transform based on normalizing flows for lossy image compression with improved performance. In addition, advanced normalizing flows can also have wide applications for lossless compression, such as integer normalizing flows \cite{hoogeboom2019integer,berg2020idf++}. While we focus on lossy compression in general, the entropy coding module in the lossy compression framework is basically a lossless compression module. 

In summary, the aforementioned four categories of probabilistic generative models contribute to both the theoretical foundations and the technical improvements for neural image compression, fostering significant progress in this domain. 

\subsubsection{Generative Image Coding for Perceptual Quality}

\tcb{It is known that objective metrics such as PSNR have discrepancies with human perception \cite{blau2018perception}.} Therefore, more and more researchers start to pay attention to the research of image coding for perceptual quality. 
In addition to the success of neural image compression methods based on variational autoencoders (VAE) and other probabilistic generative models, recent advances in image compression have seen promising results through the combination of generative models and compression models, especially by combining compression models with GANs and diffusion models \cite{ho2020denoising}. These approaches provide different strategies to achieve better rate-distortion performance and address concerns about perceptual quality without significantly increasing complexity.

GANs consist of a generator and a discriminator network. These networks are trained adversarially. In the context of image compression, GANs have been employed to generate realistic and high-quality images, while simultaneously optimizing compression efficiency. The generator learns to produce compressed representations of images, and the discriminator evaluates the authenticity of the generated images, creating a dynamic interplay that enhances both the visual fidelity and the compression effectiveness. Agustsson et al. \cite{agustsson2019generative} propose a GAN-based image compression architecture operating at extremely low bitrates, which can synthesize hard-to-store details and reduce artifacts. In addition, semantic maps can be useful for efficiently synthesizing less significant areas, if they are accessible. HiFiC \cite{mentzer2020high} investigates novel structures such as conditional GANs and normalization layers to preserve fidelity while generating visually pleasing results. Multi-Realism \cite{agustsson2023multi} designs a model with distortion-realism trade-off, allowing users to control the level of detail in reconstructed images.

Most recently, combining diffusion models and compression models has gained increasing attention. Unlike building a new lossy image compression framework with diffusion models as mentioned in Section \ref{sec:framework_genertive_model}, such a combination can achieve a good balance between complexity and perceptual quality. 
Yang et al.~\cite{yang2023lossy} propose a novel image compression framework which applies a VAE-style encoder to map images to latent variables and a diffusion model as the decoder conditioned on quantized latents. DIRAC~\cite{ghouse2023residual} leverages a diffusion model to enhance perceptual quality by producing residuals conditioned on an initially reconstructed image, which is decoded by an image codec with minimal distortion. Furthermore, HFD~\cite{hoogeboom2023highfidelity} explores an advanced noise schedule and sampling procedure and designs a patch-wise approach for high-resolution image reconstruction. All these works effectively showcase the potential of integrating generative models with compression models to enhance the visual quality of decoded images. 

\tcb{Due to the different methods used for perceptual optimization, the learned codecs exhibit different distortion types than traditional codecs. The distortions of traditional codecs generally include blocking artifacts, blurring, aliasing, etc., while learned codecs may produce smoothing, generated noise, pseudo-texture and other types of distortions, which may vary depending on the optimization method and structure.}

The substantial differences in perceptual quality among different codecs therefore require more demanding criteria for the study of accurate and explainable visual quality assessment metrics, which is also key to codec optimization. To this end, the MPEG Visual Quality Assessment ad-hoc group (AG 5) has been working to maintain a dataset of Compressed Video for study of Quality Metrics (CVQM). During the 144th MPEG meeting, MPEG AG 5 issued a call \cite{mpegcall} for learning-based video codecs for the study of quality assessment. This is because MPEG anticipates that the reconstructed videos compressed with learning-based codecs will have different types of distortions compared to those produced by the traditional block-based motion-compensated video coding. MPEG will consider inviting responses that meet the call's requirements to submit compressed bitstreams for further study and potential inclusion into the CVQM dataset.

\subsubsection{The Rate-Distortion-Perception Trade-off}

The theoretical foundations of lossy compression in mathematics are rooted in Shannon’s seminal work on the rate-distortion theory \cite{shannon1959coding}, where the distortion term is usually measured by objective metrics such as PSNR. However, in recent years, it has become increasingly accepted that 'low distortion' is not a synonym for 'high perceptual quality'. In fact, optimizing one often comes at the expense of the other \cite{blau2019rethinking}. Following the mathematical notion of perceptual quality in \cite{blau2018perception}, \tcb{the perfect perceptual quality is achieved when}
\begin{equation}
\label{equation1}
p_{\vphantom{\hat{X}}{X}} = p_{\hat{X}},
\end{equation}
where the input image is $X$, the decoded image is $\hat{X}$, and $p_X$ is the distribution of the input images. 
On the basis of this, the work of \cite{blau2019rethinking} provides a systematic study of the rate-distortion-perception trade-off. An important conclusion is later discovered by the authors in \cite{yan2021perceptual,zhang2021universal} who assert that ``We proved that, for fixed bit rate, the cost of imposing a perfect perception constraint is exactly a doubling of the lowest achievable MSE." In other words, the PSNR of the decoded image with perfect perceptual quality is 3dB lower than the PSNR of the decoded image with the smallest distortion at the same bitrate. This discipline provides an insightful guide for neural image compression models targeting perceptual quality.

In addition, Freirich et al. \cite{freirich2021theory} establish the achievable distortion perception region and provide a geometric interpretation of the optimal interpolator in Wasserstein space. Chen et al. \cite{chen2022rate} study the subtle differences between the weak- and strong-sense definitions of perceptual quality and analyze the role of randomness in encoding and decoding. While most of these works start their analyses with a toy example, the work of \cite{anonymous2023idempotence} successfully applies pre-trained unconditional generative models to real-world images and is able to achieve better perceptual quality as established in Eq. \eqref{equation1} by proposing to pursue the idempotence in neural image compression. 

In short, the rate-distortion-perception trade-off is attracting increasing attention in both theories and applications to practical problems. We believe that research on such a trade-off will continuously contribute to better approaches for generative image coding. 

\begin{figure*}[t!]
\centering
\resizebox {1\linewidth} {!} {
\begin{tikzpicture}
\draw [decorate,decoration={brace,amplitude=5pt,raise=4ex}]
  (0,0.5) -- (6.5,0.5) node[midway,yshift=4em]{\large \textbf{Autoencoder based Coding Model}};

  \node[align=left] at (3.5,-0.5) {\small Habibian2019 (ICCV 2019) \cite{habibian2019video}\\
  \footnotesize{\textcolor{gray}{Video compression with rate-distortion autoencoders}}\\ \small Liu2020 (ECCV 2020) \cite{liu2020conditional}\\
  \footnotesize{\textcolor{gray}{Conditional entropy coding for efficient video compression}}   \\  \small VCT (NeurIPS 2022) \cite{mentzer2022vct}\\ \footnotesize{\textcolor{gray}{VCT: A video compression transformer
  }}};
  
  \draw [decorate,decoration={brace,amplitude=5pt,raise=4ex}]
  (4.5,-3.5) -- (16,-3.5) node[midway,yshift=4em]{\large \textbf{Hybrid Coding Model}};
  \draw [decorate,decoration={brace,amplitude=5pt,raise=4ex}]
  (0,-5) -- (9,-5) node[midway,yshift=4em]{Explicit Residual Coding};
  \draw [decorate,decoration={brace,amplitude=5pt,raise=4ex}]
  (11.5,-5) -- (20,-5) node[midway,yshift=4em]{Implicit Residual Coding};
  \node[draw,align=left] at (15,1) {\Huge Neural Video Compression Models};

\node[align=left] at (5.3,-10) {
\textbf{Unidirectional Reference Coding:}\\
\small PMCNN (TCSVT 2019) \cite{chen2019learning}\\
\footnotesize{\textcolor{gray}{Learning for video compression}}\\ 
\small Rippel2019 (ICCV 2019) \cite{rippel2019learned}\\
\footnotesize{\textcolor{gray}{Learned video compression}}\\
\small DVC (CVPR 2019) \cite{lu2019dvc}\\
\footnotesize{\textcolor{gray}{DVC: An end-to-end deep video compression Framework}}\\
\small RaFC (ECCV 2020) \cite{hu2020improving}\\
\footnotesize{\textcolor{gray}{Improving deep video compression by resolution-adaptive flow coding}}\\
\small SSF (CVPR 2020) \cite{agustsson2020scale}\\
\footnotesize{\textcolor{gray}{Scale-space flow for end-to-end optimized video compression}}\\
\small M-LVC (CVPR 2020) \cite{lin2020m}\\
\footnotesize{\textcolor{gray}{M-LVC: multiple frames prediction for learned video compression}}\\
\small FVC (CVPR 2021) \cite{hu2021fvc}\\
\footnotesize{\textcolor{gray}{FVC: A new framework towards deep video compression in feature space}}\\
\small DVCPro (TPAMI 2021) \cite{lu2020end}\\
\footnotesize{\textcolor{gray}{An end-to-end learning framework for video compression}}\\
\small RLVC (JSTSP 2021) \cite{yang2020learningRLVC}\\
\footnotesize{\textcolor{gray}{Learning for video compression with recurrent auto-Encoder and recurrent probability}}\\
\small C2F (CVPR 2022) \cite{hu2022coarse}\\
\footnotesize{\textcolor{gray}{Coarse-to-fine deep video coding with hyperprior-guided mode prediction}}\\
\small ENVC (TIP 2023) \cite{guo2023learning2}\\
\footnotesize{\textcolor{gray}{Learning cross-scale weighted prediction for efficient neural video compression}}\\
};

\node[align=left] at (5.5,-18.5) {
\textbf{Bidirectional Reference Coding:}\\
\small Wu2018 (ECCV 2018) \cite{wu2018video}\\
\footnotesize{\textcolor{gray}{Video compression through image interpolation}}\\ 
\small Djelouah2019 (ICCV 2019) \cite{djelouah2019neural}\\
\footnotesize{\textcolor{gray}{Neural inter-frame compression for video coding}}\\ 
\small Cheng2019 (CVPR 2019) \cite{cheng2019learning}\\
\footnotesize{\textcolor{gray}{Learning image and video compression through spatial-temporal energy compaction}}\\ 
\small HLVC (CVPR 2020) \cite{yang2020learning}\\
\footnotesize{\textcolor{gray}{Learning for video compression with hierarchical quality and recurrent enhancement}}\\ 
\small Pourreza2021 (ICCV 2021) \cite{Pourreza_2021_ICCV}\\
\footnotesize{\textcolor{gray}{Extending neural P-frame codecs for B-frame coding}}\\ 
\small LHBDC (TIP 2021) \cite{yilmaz2021end}\\
\footnotesize{\textcolor{gray}{End-to-end rate-distortion optimized learned hierarchical bi-directional video compression}}\\ 
\small ALVC (TCSVT 2022) \cite{yang2022advancing}\\
\footnotesize{\textcolor{gray}{Advancing learned video compression with in-loop frame prediction}}\\ 
};

\node[align=left] at (17.8,-10.5) {
\textbf{Unidirectional Reference Coding:}\\
\small ModeNet (MLSP 2020) \cite{ladune2020modenet}\\
\footnotesize{\textcolor{gray}{ModeNet: mode selection network for learned video coding}}\\ 
\small Ladune2020 (MMSP 2020) \cite{ladune2020optical}\\
\footnotesize{\textcolor{gray}{Optical flow and mode selection for learning-based video coding}}\\ 
\small ELF-VC (ICCV 2021) \cite{rippel2021elf}\\
\footnotesize{\textcolor{gray}{ELF-VC: efficient learned flexible-rate video coding}}\\ 
\small DCVC (NeurIPS 2021) \cite{li2021deep}\\
\footnotesize{\textcolor{gray}{Deep contextual video compression}}\\ 
\small Brand2022 (PCS 2022) \cite{brand2022benefits}\\
\footnotesize{\textcolor{gray}{On benefits and challenges of conditional interframe video coding in light of information theory}}\\ 
\small DCVC-TCM (TMM 2022) \cite{sheng2022temporal}\\
\footnotesize{\textcolor{gray}{Temporal context mining for learned video compression}}\\ 
\small CANF-VC (ECCV 2022) \cite{ho2022canf}\\
\footnotesize{\textcolor{gray}{CANF-VC: conditional augmented normalizing flows for video compression}}\\ 
\small DCVC-HEM (ACMMM 2022) \cite{li2022hybrid}\\
\footnotesize{\textcolor{gray}{Hybrid spatial-temporal entropy modelling for neural video compression}}\\ 
\small MIMT (ICLR 2023) \cite{xiang2022mimt}\\
\footnotesize{\textcolor{gray}{MIMT: masked image modeling transformer for video compression}}\\ 
\small DCVC-MIP (CVPR 2023) \cite{qi2023motion}\\
\footnotesize{\textcolor{gray}{Motion information propagation for neural video compression}}\\ 
\small DCVC-DC (CVPR 2023) \cite{li2023neural}\\
\footnotesize{\textcolor{gray}{Neural video compression with diverse context}}\\ 
\small Brand2024 (TCSVT 2024) \cite{brand2024conditional}\\
\footnotesize{\textcolor{gray}{Conditional Residual Coding: A Remedy for Bottleneck Problems in Conditional Inter Frame Coding}}\\ 
};

\node[align=left] at (16.7,-17.4) {
\textbf{Bidirectional Reference Coding:}\\
\small B-CANF (TCSVT 2023) \cite{chen2023b}\\
\footnotesize{\textcolor{gray}{B-CANF: adaptive B-frame coding with conditional augmented normalizing flows}}\\ 
\small TLZMC (CVPR 2023) \cite{alexandre2023hierarchical}\\
\footnotesize{\textcolor{gray}{Hierarchical B-frame video coding using two-Layer CANF without motion coding}}\\ 
};

\end{tikzpicture}}
    \caption{Neural Video Compression models.}
    \label{VideoComModels}
\end{figure*}

\subsection{Video Coding with Generative Models}

While the field of neural image compression has been fully developed, the field of neural video coding is also experiencing tremendous development. In this section, we discuss the neural video coding frameworks from two categories, as depicted in Fig. \ref{VideoComModels}: the autoencoder-based coding models and hybrid coding models. Moreover, we analyze the development of generative video coding for perceptual quality improvement.

\subsubsection{Autoencoder-based Coding Models for Video Compression}

A few works regard neural video compression as an extension of the autoencoder-based image compression framework, where the coding pipeline is generally divided into two parts: transform coding and conditional entropy coding. Specifically, a 3D or 2D autoencoder is used to transform the video sequence into quantized features to encode, and then a conditional entropy coder that combines spatiotemporal information is used for entropy coding \cite{habibian2019video, liu2020conditional, mentzer2022vct}. Given the independence of distortion introduced in the time domain, this framework effectively avoids issues like error propagation. However, the algorithm's overall complexity tends to be high. Redundancy removal primarily occurs in the entropy coding module, which does not fully capitalize on the benefits of transform coding.

\subsubsection{Hybrid Coding Models for Video Compression}
The current mainstream neural video compression framework still uses a hybrid coding framework that combines inter-frame motion estimation, which is similar to the traditional video coding framework. Generally, the coded motion information and the previous decoded frame are used to generate the reference frame, while the residual information between the current frame and the reference frame is encoded at a later stage.

An \tcb{early-stage study \cite{chen2019learning} conducted in early 2018 introduced a block-based hybrid generative module called PixelMotionCNN to model spatiotemporal coherence; the module utilizes effectively predictive coding together with additional components of iterative analysis/synthesis to reach comparable compression results with an H.264 codec.} Lu et al. \cite{lu2019dvc, lu2020end} proposed to replace every module in a traditional video compression framework with neural networks. In particular, the estimated optical flow is treated as motion information and encoded together with the residual frame by two different autoencoders. At the same time, Rippel et al. \cite{rippel2019learned} also proposed a highly complete model based on similar ideas, which not only replaced the traditional coding module, but also expanded it into a more general model that compresses the generalized state. Since then, numerous studies have been dedicated to exploring methods for acquiring effective representation and precise modeling of motion information. Some works attempt to perform hierarchical processing in high-dimensional representation space to achieve a better rate-distortion balance \cite{hu2021fvc, hu2022coarse}, while other works explore better motion estimation and compensation \cite{agustsson2020scale, lin2020m, guo2023learning}, motion compression \cite{hu2020improving} or time correlation mining \cite{yang2020learningRLVC} to improve the accuracy of representation. In addition, some works start from the reference relationship between frames and use frame interpolation models, recurrent neural networks, etc. to explore the value of bidirectional reference relationships \cite{wu2018video, djelouah2019neural, cheng2019learning, yang2020learning, Pourreza_2021_ICCV, yilmaz2021end, yang2022advancing}. 

\tcb{Furthermore, by utilizing a hybrid coding framework in conjunction with motion estimation, a different set of studies have aimed to eliminate the need for residual frames. Instead, these studies focus on utilizing the information obtained from motion compensation results to enhance the encoding and decoding process of the frame being encoded.} The former can be considered as explicit residual coding, while the latter can be considered as implicit conditional coding. \tcb{Theoretically, as was demonstrated and discussed in \cite{brand2022benefits}, the latter has a higher rate-distortion upper bound.} \tcb{In order to introduce the skip mode structure, Ladune et al. \cite{ladune2020modenet} firstly built and tried out the concept of conditional coding in the area of learned video compression, and developed their architecture with an optical flow network in \cite{ladune2020optical}.} While \cite{rippel2021elf} introduced \tcb{a more general design} of conditional coding from an engineering perspective, \cite{li2021deep} further \tcb{summarized the core idea} and designed an efficient coding model based on it. The follow-up work focuses on more efficient information transmission \cite{sheng2022temporal} and functional improvements such as supporting variable rates \cite{li2022hybrid}. Several following works are put forward through architecture and module improvements \cite{ho2022canf, xiang2022mimt, qi2023motion, li2023neural}. The later work \cite{brand2024conditional} re-added the residual structure in conditional coding to solve the possible bottleneck problem. As the representative work of implicit conditional coding, the work \cite{li2023neural} has been able to surpass the low-delay configuration of the H.266/VVC standard reference software VTM in terms of both the objective metrics of RGB PSNR and MS-SSIM.

\subsubsection{Generative Video Coding for Perceptual Quality}

Although most neural video compression works still strive to improve objective performance with metrics such as PSNR and MS-SSIM, a few works have drawn inspiration from image compression techniques to enhance the visual quality of compressed videos. Yang et al. \cite{DBLP:conf/ijcai/YangTG22} investigated perceptual optimized video compression with recurrent conditional GAN. Mentzer et al. \cite{mentzer2022neural} directly explored conditional GAN training and verified the effectiveness of this method through user studies, while another work analyzes perception loss functions for learned video compression \cite{phan2023choice}. 
The work \cite{yang2023diffusion} presents a diffusion probabilistic modeling approach for video generation, drawing inspiration from recent advances in neural video compression, which has the potential to offer valuable insights for enhancing the perceptual performance of video coding.
\begin{figure*}
  \centering
  \centerline{\includegraphics[width=16cm]{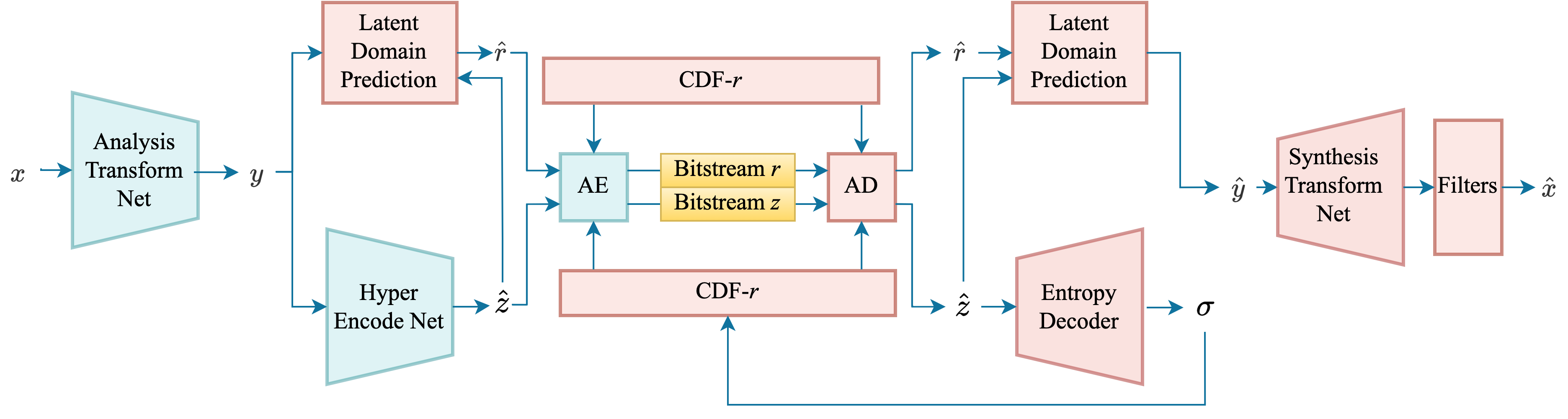}} 
\caption{VAE-based JPEG-AI framework. $x$ and $\hat{x}$ denote the original input image and the reconstructed images, respectively. The red modules are standardized in JPEG-AI. The \textcolor{black}{blue-green} modules are the encoder side operations~\cite{overview_slide}.
}
\label{jpeg_ai_framework}
\end{figure*}

\begin{figure}[t]
    \centering
    {\includegraphics[width=4.3cm]{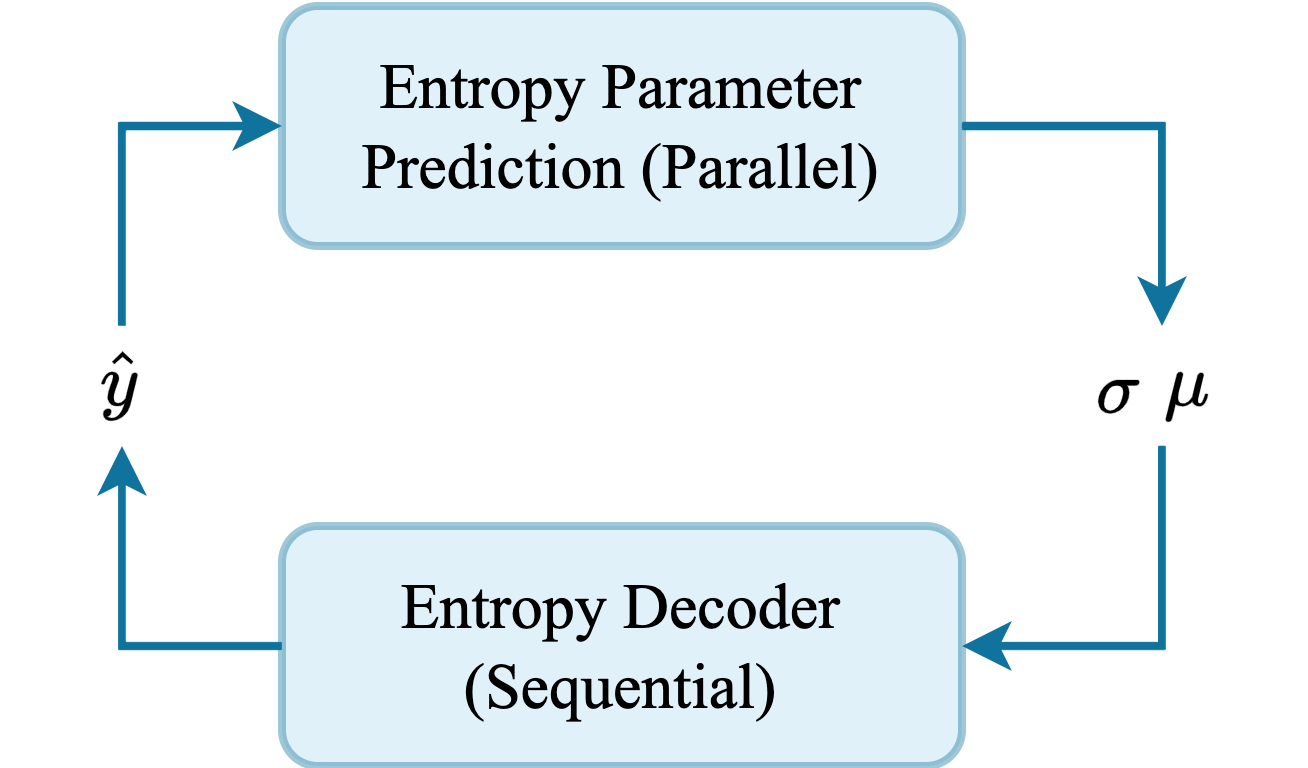}}%
    \hfill
    {\includegraphics[width=4.3cm]{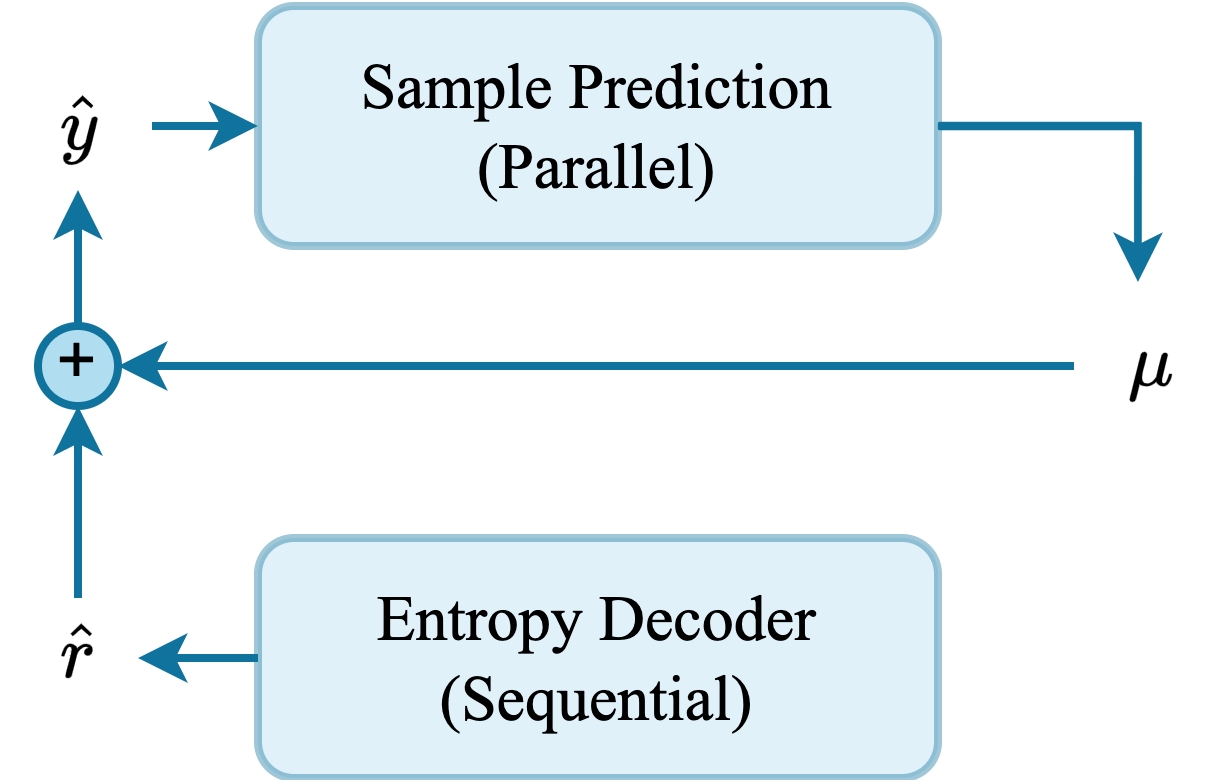}}%
    \caption{\color{black}Illustration of the sequential architecture and the decoupled architecture; left: sequential architecture; right: decoupled architecture~\cite{decoupled}~\cite{CFP_bytedance}.}
     \label{decoupled}
\end{figure}

\begin{figure}[t]
  \centering
  \centerline{\includegraphics[width=9cm]{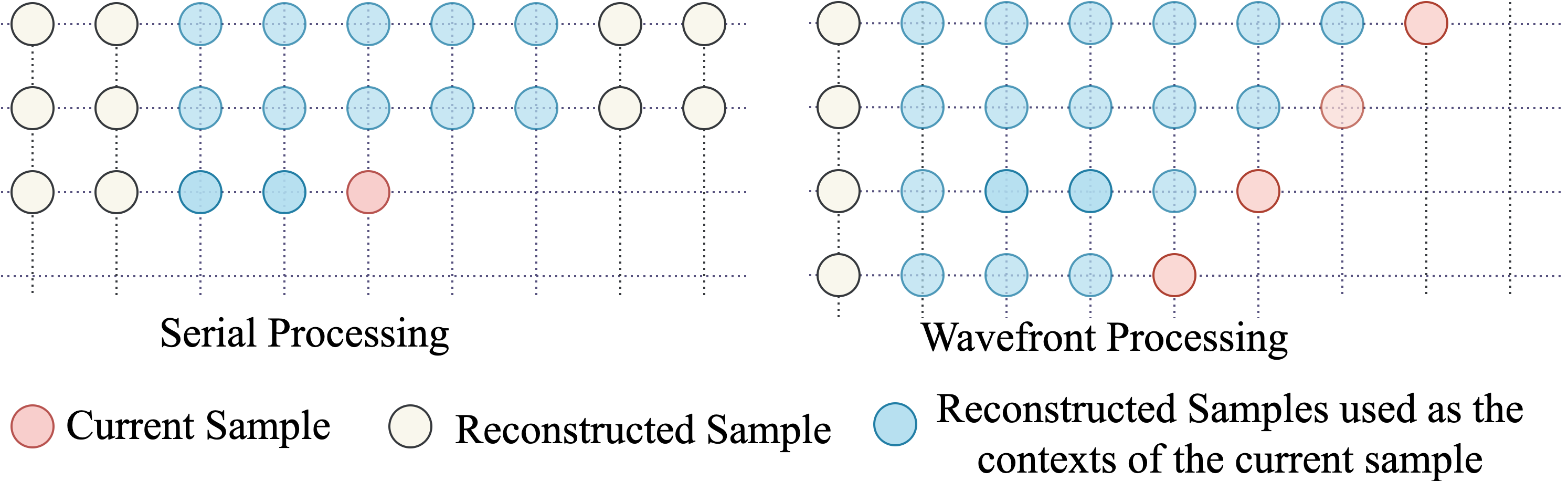}} 
\caption{Illustration of the serial processing based on the raster-scan order and wavefront processing~\cite{decoupled,CFP_bytedance}.
}
\label{wavefront}
\end{figure}

\begin{figure}[t]
  \centering
  \centerline{\includegraphics[width=9cm]{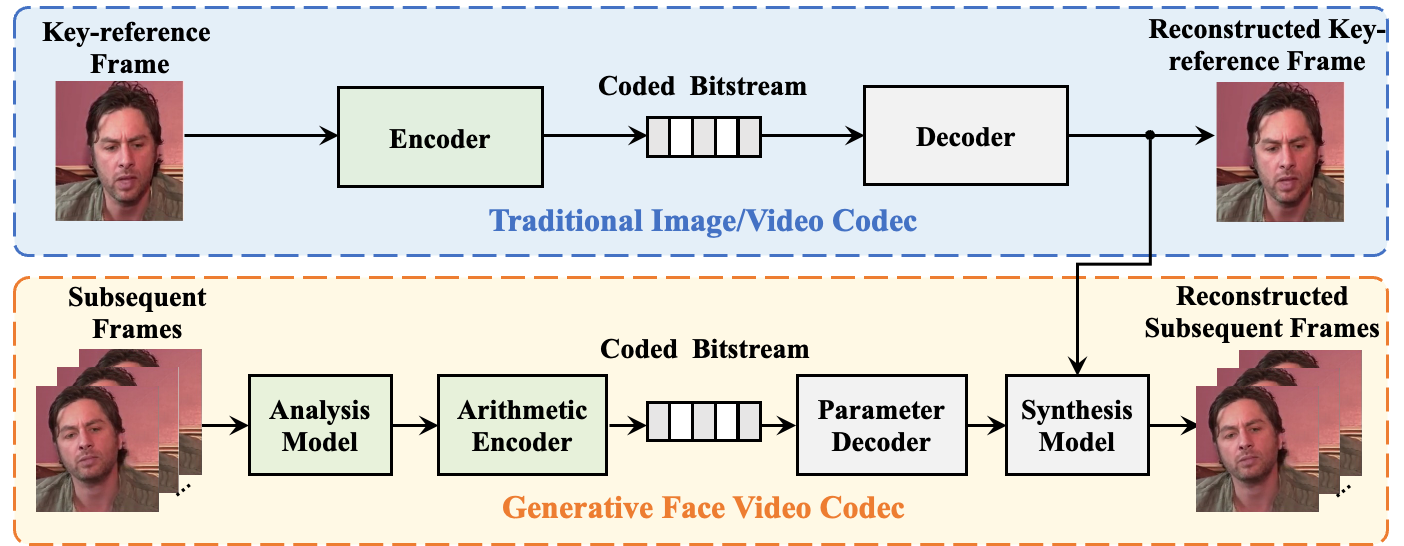}} 
\caption{Framework of Generative Face Video Coding.}
\label{gfvc}
\end{figure}

\section{{\color{black}Standardization Activities} related to visual signal coding with generative models}
\label{sec:Stan}
This section briefly overviews the standardization activities related to visual signal coding with generative models. Specifically, the progress of image and video coding using generative models is described in the following subsections.

\subsection{JPEG AI Standardization Activities}

Traditional image coding schemes, such as JPEG~\cite{wallace1992jpeg}, JPEG2000~\cite{skodras2001jpeg} and intra coding in {\color{black}H.264/AVC}~\cite{wiegand2003overview}, {\color{black}H.265/HEVC}~\cite{sullivan2012overview}, AV1~\cite{AV1} and {\color{black}H.266/VVC}~\cite{bross2021overview}, are developed in a classical paradigm, including block partition~\cite{kim2012block}, intra prediction~\cite{pfaff2021intra}, transformation~\cite{zhao2021transform}, and entropy coding~\cite{schwarz2021quantization}.
Recently, learned image compression methods~\cite{balle2018variational,cheng2020learned,minnen2018joint} have achieved significant progress and superior improvements in rate-distortion performance, attracting lots of attention from both industry and academia.
One attractive feature lies in that learned image coding schemes are able to adapt to different applications with moderate additional effort. They can achieve good compression results for machine vision and image processing tasks by changing the optimization targets~\cite{ascenso2023jpeg}.

JPEG AI, a learning-based image coding standard, was under development~\cite{ascenso2023jpeg} at the time of writing and was expected to be finalized in 2024. JPEG AI is the first AI-based image coding standard developed jointly by the ISO/IEC and ITU-T. 
The scope of JPEG AI is the creation of a learned image coding standard aimed at offering a royalty-free baseline that achieves significantly better compression performance than existing image coding standards and meanwhile provides enhanced performance for image processing and computer vision tasks. JPEG AI is designed to cater to versatile real-world image applications, such as surveillance, cloud and edge storage, autonomous driving, and visual data transmission and distribution. The Working Draft (WD) and the Committee Draft (CD) of the JPEG AI core coding system were released in 2023. The international standard is expected to be published in 2024. 

Analogously to many existing learning-based image coding schemes, JPEG AI adopts a VAE-based framework~\cite{balle2018variational,minnen2018joint}. The input image is compactly converted into its latent representation by an analysis transform network at the encoder side. The synthesis transform network recovers the input image from the latent representation at the decoder side. Moreover, the latent representation is entropy coded as a bitstream. The framework of JPEG AI is illustrated in Fig.~\ref{jpeg_ai_framework}, {\color{black}where only the modules in red color are standardized}.
The overall framework is optimized in an end-to-end way by minimizing the rate-distortion cost, where the distortion term can be calculated with the Mean Squared Error (MSE) or Multi-Scale Structural SIMilarity (MS-SSIM)~\cite{wang2003multiscale} for optimized objective quality during training. Generative models such as GAN have been trialed for optimizing perceptual quality during the development of JPEG AI~\cite{decoupled}. In particular, the GAN-based perceptual loss~\cite{lucas2019generative} is introduced as one of the distortion terms for training the networks, generating perceptual-oriented models that provide better visual quality in low-rate coding scenarios. 

JPEG AI Common Test and Training Conditions (CTTC)~\cite{cttc} present guidelines on the training and performance evaluation of the JPEG AI Verification Model (VM). The training dataset includes 5000 images, and the validation dataset contains 350 images. The test dataset has 50 images. Various quality metrics, including MS-SSIM, Video Multimethod Assessment Fusion (VMAF) metric~\cite{rassool2017vmaf}, Visual Information Fidelity (VIF)~\cite{ponomarenko2007between}, are employed to evaluate the coding distortion of the reconstructed image, emphasizing the perceptual quality of human vision. The target bit-rates are \{0.06, 0.12, 0.25, 0.50\} bits-per-pixel (bpp). The coding complexity is measured in the number of multiply-accumulate operations per pixel (kMAC/pxl).

Compared with the VVC intra coding~\cite{VTM11}, JPEG AI VM-4.3 achieves a 28.5\% BD-Rate saving at the high operation point. At the base operation point, the averaged coding gain is  16.4\%~\cite{VM4_software}. 

\subsubsection{VAE-based Transformation}
The analysis and synthesis transform networks of JPEG AI are built with residual blocks, attention blocks~\cite{zhang2019residual}, and activation layers for nonlinear conversion. These networks are operated with the YUV 420 color format by default, such that a color space conversion layer is provided at the beginning of the analysis transform at the encoder side.
Moreover, the luma and chroma components employ separated analysis and synthesis transform networks to reduce peak memory usage. The network is deeper and heavier for the primary component (luma) to better compress and reconstruct the texture details. As such, given the input image, the analysis transform cooperating with latent domain prediction tools maps the image into its latent representations, which include the residual, prediction and entropy information. The synthesis transform generates the visual signals based on the latent representations parsed from the bitstream. The hyper-prior networks, including the hyper encoder network, hyper decoder network and hyper-scale decoder network, are also built on top of the VAE, where the distribution parameters are embedded as the hyper-prior term for high-efficiency entropy coding.

During the standardization of JPEG AI, two operation points, namely, the base operation point and the high operation point, are developed to cater to different application scenarios.
The layer design such as the convolution kernel size and the upscaling strategy of these two operation points are different. The decoding complexities of the base operation point and the high operation point are around 20 kMAC/pxl and 200 kMAC/pxl, respectively~\cite{VM4_software}.

In particular, the base operation point is with minimal networks for encoding and decoding, providing a lightweight decoder suitable for the deployment on mobile devices. The coding tools such as the residual and variance scale (RVS), latent scale before synthesis, and enhancement filters are all disabled in the configuration of the base operation point.
At the high operation point, attention models, including the Transformer-based Attention Module (TAM) and the Convolutions-based Attention Block (CAB), are enabled in the synthesis transform network to enhance the generative capability of the decoder~\cite{overview_slide}. 
The high operation point enables all the coding tools for enhanced compression performance.

\subsubsection{AR-based Context Modeling and Entropy Coding}

A decoupled architecture~\cite{decoupled,CFP_bytedance} is designed for entropy coding to decouple the decoding dependencies between entropy decoding and latent sample reconstruction, as illustrated in Fig.~\ref{decoupled}. 
Unlike the previous design, where the arithmetic decoder is interleaved with the generation of the entropy parameters, the entropy decoding process in the decoupled architecture is independent from the latent sample reconstruction, leading to significant reductions in decoding time. This design philosophy has been adopted by traditional video coding standards. 
To be more specific, a hyper decoder and a hyper-scale decoder are involved in the decoupled architecture. The prediction of latent samples is reconstructed from the hyper decoder, and the Gaussian variance is recovered from the hyper-scale decoder. After the quantized residual samples are obtained from the bitstream, the reconstruction process is invoked. 
In this way, the latent reconstruction and the entropy decoding are separated; consequently, the entropy decoding will not be suspended by the latent reconstruction.

Autoregressive models exhibit prominent context modeling capability, which has been widely employed in the learned image compression. Following the raster-scan sequential processing order, the reconstruction of the current sample relies on the previous neighboring reconstructed samples.
As such, the main issue of the autoregressive model lies in the strict sequential processing during context modeling, which hinders its deployment in real-world applications. To facilitate parallelization and improve GPU utilization efficiency, the wavefront parallel strategy is supported in JPEG AI context modeling~\cite{decoupled,CFP_bytedance}, with which the latent can be constructed in the row-wise concurrent processing order, as shown in Fig.~\ref{wavefront}. The context network involves neighboring reconstructed samples as input and yields multiple outputs simultaneously, leading to much reduced decoding time.

Even though the wavefront decoding strategy can enhance the parallelization of the context modeling and latent prediction, there is still room for further improvements to the prediction speed. A Multi-stage Context Modeling (MCM) method~\cite{MCM} is adopted in JPEG AI, which enhances parallelization to save decoding time and maintain coding performance. More specifically, the MCM is built on top of the decoupled architecture, which is used as the replacement of the context model when predicting the mean of the latent representation, so as to further reduce the decoding complexity of the entropy coding module. Instead of sequentially modeling the contexts, the tensor of the latent representation is divided into 8 groups through down-shuffle, and sub-groups are concatenated in the channel dimension. The context modeling process can be regarded as the implicit prediction of the element in the latent representation with the reconstructed group elements. MCM achieves 97\% speedup during the latent sample prediction in the entropy decoder, with only a 2.1\% BD-Rate loss. 

\subsection{Exploration of Neural Network-based Video Coding}

Dating back to \textcolor{black}{June 2020}, the Joint Video Exploration Team (JVET) of ITU-T VCEG and ISO/IEC MPEG established an “ad hoc group” (AhG) on NNVC at the 130th MPEG meeting and 19th JVET meeting~\cite{JVET-S2000}. Since then, two categories of NNVC have been extensively studied within this AhG. 
{\color{black}In the first category, the neural network-based (NN-based) modules are embedded in the traditional video coding framework to improve coding performance. In particular, the NN-based modules are used to replace the classical coding modules such as intra prediction~\cite{intraJVET-T0073, Dumas2023Neural,intraNNAB0149, intraJVET-AD0212}, inter prediction~\cite{interJVET-AC0090, interJVET-AC0114}, in-loop filter~\cite{inloopJVET-T0079,inloopJVET-T0088, inloopJVET-AC0089, lopJVET-AD0156, lopJVET-AF0043}, post filter~\cite{postJVET-V0075} and resampler \cite{srJVET-T0096,srJVET-AC0196}. The overall framework is optimized with rate-distortion criteria following the traditional coding philosophy.
In the second category, the coding paradigm is achieved with the full neural network. More specifically, predictive coding is explored, which employs optical flows to realize the inter prediction. Then, residual signals are derived and entropy coded with an autoencoder. Conditional coding is also investigated in the full neural network based video coding, wherein the prediction is achieved by deriving the latent representation with the autoencoder~\cite{JVET-AA0063, e2eJVET-AC0091}. 

It is noted that the proposed end-to-end learned coding method in \cite{JVET-AA0063} is only applied to inter frames. For coding intra frames (i.e., I frame), the traditional video coding standards, such as H.266/VVC intra coding or BPG compression, are applied. Meanwhile, a common test model, also knowned as Neural Network-based Video Coding (NNVC), was initially produced with two NN-based in-loop filtering tools \textcolor{black}{in July 2022} and maintained for the exploration experiments on NN-based technologies. NNVC evolves with the investigation activities of each meeting cycle. 
As of Jan. 2024, NNVC-7.1 comprises two main NN-based modules including the NN-based intra prediction and NN-based in-loop filtering, with the goal of enhancing the coding performance of the traditional tools within the current H.266/VVC standard.
To be more specific, with the NN-based intra prediction, fully connected neural networks are utilized to establish a nonlinear mapping between neighboring reference samples and samples in the current block~\cite{Dumas2023Neural, dumas2020iterative}.
Moreover, the NN-based module additionally produces auxiliary outputs that assist in constructing the Most Probable Mode (MPM) list and selecting transform kernels for subsequent processes. Regarding the NN-based in-loop filtering, a convolutional neural network-based in-loop filter is utilized to enhance the reconstruction quality and recovering capability~\cite{Li2022Deepfixedpoint,Li2022rdo,Liu2023combined,Li2023additional,li2023idam}.
The NN-based filter undergoes iterative training to tackle the problem of excessive filtering, as outlined in \cite{li2023idam}.  
Enhancing performance involves additional considerations such as leveraging coded information, selecting parameters, adapting inference granularity, scaling residuals, incorporating temporal filtering, integrating deblocking filtering, aligning with Rate-Distortion Optimization (RDO).}
In addition, the deep filter supports two trade-off points in terms of compression complexity and efficiency, namely the low operation point (LOP) at 17kMAC/pixel and high operation point (HOP) at 477 kMAC/pixel.
 \textcolor{black}{The coding performance of the NN-based filter and intra prediction in NNVC-7.1 as compared to the H.266/VVC reference software VTM-11.0 is summarized in Table \ref{table_NNVC}~\cite{JVET-AG0014}. From this comparison, it is evident that significant BD-rate gains of up to -13.59\% and -12.55\% are observed for the Y component under Random Access (RA) and All Intra (AI) configurations, respectively. This underscores the considerable potential of neural network-based coding tools in advancing video compression performance.}

\begin{table}[!t]
\renewcommand{\arraystretch}{0.7}
\caption{\color{black}Coding Performance (BD-Rate) of NNVC-7.1 over H.266/VVC Reference Software VTM-11.0~\cite{JVET-AG0014}.}
\label{table_NNVC}
\centering
\tabcolsep0.01in
\begin{tabular}{c|ccc|ccc}
\toprule
\multirow{2}{*}{Methods} &\multicolumn{3}{c|}{AI} &\multicolumn{3}{c}{RA} \\
        \cmidrule{2-7}
                &Y        &U &V          &Y         &U &V   \\  
\midrule
LOP+NN{-}Based Intra       &-8.13\%  & -13.28\% & -13.42\%   &-6.89\%   &-13.17\%  &-12.15\%  \\ 
\midrule
HOP+NN{-}Based Intra     &-12.55\%  & -11.37\% & -13.05\%   & -13.59\%  &-12.47\%      & -14.18\%  \\ 
\bottomrule
\end{tabular}
\end{table}

\subsection{Exploration of Generative Face Video Coding}

Different from early model-based coding (MBC) techniques~\cite{7268565,1457470,lopez1995head,150969}, generative face video coding (GFVC) schemes ~\cite{facebook2021,ultralow,9859867,CHEN2022DCC,icip2022zhao,chen2023csvt,chen2023interactive,YIN2024DCC} exploit the excellent generative ability of deep generative models~\cite{goodfellow2014generative,NEURIPS2021_49ad23d1} to improve the face reconstruction quality and realize ultra-low bitrate face video communications. Specifically, the encoder employs the traditional image/video codec to compress the key-reference frames of a face video, and encodes the subsequent inter frames into compact transmitted symbols (\textit{e.g.,} landmarks/keypoints, compact feature and facial semantics). Besides, the decoder feds these decoded key-reference frames and compact facial representations into the deep generative model to learn the temporal evolution and reconstruct these face frames. The typical framework of GFVC is depicted in Fig. \ref{gfvc}.

Inspired by such promising rate-distortion performance, some GFVC proposals have been submitted to JVET, where they explored whether GFVC's compact symbols could be inserted into an H.266/VVC bitstream as Supplemental Enhancement Information (SEI) messages. \textcolor{black}{More specifically, the generative face video (GFV) SEI message~\cite{JVET-AD0051,JVET-AE0088} allows VVC-coded pictures to be utilized as the base (key-reference) pictures and incurs only very little overhead in the compressed bitstream to signal compact facial information with different representation types, including compact temporal features, 2D keypoints, 3D keypoints, facial semantics and others. In addition, several proposals~\cite{JVET-AG0087,JVET-AE0280, JVET-AG0088} aim to enable a more common GFV SEI syntax and specify the decoder interface with the generative neural network.}

Furthermore, Ye \textit{et al.}~\cite{m64987} made the related technical requirements regarding the exploration and potential standardization of ultra-low bitrate 2D generative face video coding methods. As such, JVET experts decided to establish a new ad hoc group to conduct GFVC investigations on software implementation, experiment coordination, interoperability studies, and other related aspects. \textcolor{black}{In particular, a unified software package~\cite{JVET-AG0042} with various GFVC methods has been proposed to allow coding to be performed using the VVC Main 10 profile in Jan. 2024. Besides, the common test conditions and software reference configurations~\cite{JVET-AG2035} have also been specified for GFVC experiments. In addition, the interoperability study in feature translation~\cite{JVET-AG0048} and {\color{black}lightweight model}~\cite{JVET-AG0139} on different GFVC systems are further investigated to allow more flexible GFVC applications within acceptable performance losses. }

In short, the current GFVC standardization activities are mainly concentrated on the design of the SEI message with rich facial representations such that it warrants the service of ultra-low rate communications, user-specified animation/filtering and metaverse-related functionalities. Certainly, there exist issues and challenges for GFVC's standardization and deployment, e.g. unstable generation quality, high decoder complexity, inadequate model interpretability and inappropriate evaluation measures.

\section{Visual Signal Processing with Generative Models}
\label{sec:GVSP}

Apart from visual signal coding, another major application venue of generative models is visual signal processing. Various generative models have been applied and adapted to different image and video processing tasks, including restoration, synthesis, editing, and interpolation. This section provides a brief overview of the important works on these research topics, highlighting the important role of generative models for these applications. It is noted that there is another important research area focusing on the generation of visual signals from texts. Due to the limited length of this paper, here we solely review the approaches that take visual signals as input.

\subsection{Generative Visual Signal Restoration}

Image and video restoration is a process to recover the high-quality version of a visual signal that is associated with different perceptual artifacts. These artifacts can be generated in different production stages, including content capture, transmission, and display, which could potentially degrade the perceptual quality of visual signals and reduce the effectiveness of high-level computer vision algorithms (e.g., detection and classification)~\cite{anantrasirichai2022artificial}. Various restoration methods are commonly applied at different steps in the production pipeline. According to the nature of the artifacts,  restoration methods can be classified as denoising (to remove camera and production noises), deblurring (to reduce focal or motion blurs), dehazing/deraining (to alleviate the haze and raining effect), super-resolution (for spatial resolution upsampling) and compression enhancement (to remove compression artifacts)~\cite{rota2023video}. As the review of visual signal coding with generative models has already been conducted in Section \ref{sec:coding}, here we solely focus on other restoration tasks, providing a concise summary of some key works based on generative models, which are proposed for visual signal restoration. For a more comprehensive overview of the literature on this topic, the reader is referred to references including~\cite{zhai2023comprehensive,rota2023video,chauhan2023deep,li2023diffusion}.

\subsubsection{VAE-based Restoration}

Early attempts at generative restoration include ~\cite{kingma2013auto,rezende2014stochastic}, which employ vanilla VAEs to perform denoising. Their performance has been further improved by Denoising Autoencoders (DAE)~\cite{im2017denoising}, which train VAEs with noise injected into their stochastic hidden layer. Moreover, the DAE model has been enhanced through the combination with a more advanced training methodology, resulting in Denoising Adversarial Autoencoders (AAEs)~\cite{creswell2018denoising}, and by incorporating explicit models of the image noise distribution in the decoder, with DivNoising~\cite{prakash2020fully} as a notable work. VAE-based approaches have also been applied to deblurring, with examples including \cite{nimisha2017blur,asim2020blind}, where the networks employed consist of an autoencoder to learn the image prior and an adversarial network to discriminate blurred and clean images or their features. Moreover, VAEs have also contributed to the task of dehazing/deraining. The variational image detraining (VID)~\cite{du2020variational} method uses a conditional variational encoder (CVAE) to perform probabilistic inference that increases the diversity of prediction. pWAE (pixel-wise Wasserstein autoencoder)~\cite{kim2021pixel} introduces 2D latent tensors to the Wasserstein autoencoder to allow pixel-wise matching, which is reported to offer better dehazing performance compared to conventional autoencoders. Finally, VAEs have also advanced the development of superresolution. Important works include SR-VAE (Image Super-resolution via Variational AutoEncoders) ~\cite{liu2020photo} and VDVAE-SR (Very Deep Variational Autoencoder Super-resolution) ~\cite{chira2022image}. SR-VAE learns the conditional distribution of high resolution images providing their low resolution counterparts, which can generate super-resolution results with photorealistic visual quality and relatively low distortion~\cite{liu2020photo}. VDVAE-SR adapts a very deep VAE model to single-image super-resolution and employs a low resolution (LR) encoder to learn the image prior. This has been demonstrated to provide competitive super-resolution performance compared to the SotA at the time~\cite{chira2022image}. 

\subsubsection{GAN-based Restoration}

As one of the primary types of generative models, GANs have been widely used for visual signal restoration. Typically, these GAN-based approaches enable the generation of photorealistic details rather than simply minimizing the distortion between the output and the training targets. For the task of denoising, a number of GAN-based methods have been proposed in the context of natural and medical image denoising. For example, a GAN was trained in~\cite{chen2018image} to learn the noise distribution within noisy input images, based on which noise samples are generated from clean images and used to train a deep CNN for denoising. For this task, more advanced GAN architectures are also employed, with examples based on CycleGAN~\cite{anantrasirichai2021contextual}, StyleGAN~\cite{poirier2023robust}, and Wasserstein GAN~\cite{cha2020gan2gan}. In the research field of dehazing/deraining, researchers not only utilized existing GAN models for clean image and video recovery \cite{li2018single,zhang2019image}, but also developed customized GAN-based approaches for this purpose. A notable approach is FD-GAN, which employs a fusion discriminator to learn additional priors from frequency information - this allows the generation of more photorealistic dehazed images. Similarly, DW-GAN~\cite{fu2021dw} was developed by combining a GAN with a discrete wavelet transform to obtain excellent dehazing performance for images with a nonhomogeneous haze effect. For superresolution, a large number of methods have been proposed based on existing GAN models such as standard GANs~\cite{ledig2017photo,lin2019real}, conditional GANs~\cite{xie2018tempogan,kudo2019virtual}, patch GANs~\cite{maeda2020unpaired,wang2020enhanced}, relativistic average GANs~\cite{wang2018esrgan,ma2019perceptually,shang2020perceptual} and Wasserstein GANs~\cite{you2019ct}. One of the earliest but influential works is SRGAN~\cite{ledig2017photo}, which is arguably the first attempt focusing on perceptually inspired super-resolution. Different GAN variants have also been designed specifically for this task, with recent examples such as content-aware local GAN (CAL-GAN)~\cite{park2023content} and Generative and Controllable Face Super Resolution (GCFSR)~\cite{he2022gcfsr}.

\subsubsection{Restoration based on Diffusion Models}

In the past three years, being one of the most popular research topics in machine learning and computer vision~\cite{croitoru2023diffusion}, diffusion models have now been actively exploited in the context of image and video restoration. Although it is at a very early stage, this type of approach shows the promise in competing with classic CNN-based restoration methods and those based on other generative models. For example, denoising diffusion probabilistic models (DDPMs) have been employed for single- and multiple-weather image restoration (e.g., desnowing, deraining, and dehazing) in~\cite{ozdenizci2023restoring}. 
DDPMs have also inspired super-resolution approaches such as SR3 (Super-Resolution via Repeated Refinement)~\cite{saharia2023image} and SRDiff~\cite{li2022srdiff}. Moreover, a new Denoising Diffusion Restoration Model (DDRM) has been proposed in~\cite{kawar2022denoising} for multiple restoration tasks, including debluring, super-resolution, and inpainting. While various diffusion models have been used and developed for the restoration task, they can also be combined with other deep learning techniques to achieve improved restoration performance. One of the notable works in this category is the Implicit Diffusion Model (IDM)~\cite{gao2023implicit}, which integrates the implicit neural representation and diffusion models in the same framework - this allows the developed super-resolution model to perform continuous-resolution requirement. Despite the promising results generated by various diffusion models for the restoration task, its low computational efficiency has also been observed and considered a common drawback. Recently, efforts have been made to design light-weight diffusion-based restoration models, including Spectral Diffusion~\cite{yang2023diffusion} and DiffIR~\cite{xia2023diffir}.

\subsection{Generative Visual Signal Synthesis and Editing}

As another important research area in visual signal processing, image and video synthesis and editing focus on generating photorealistic content, or editing existing images or videos with a new style, background, or foreground~\cite{zhan2023multimodal}. In recent years, advances in generative models have made significant contributions to this research field. According to the input references, these synthesis and editing methods can be classified as visual guidance, audio guidance, or text guidance. Due to the limited space in this survey paper, here we mainly focus on the review of generative synthesis and editing approaches based on visual guidance. For a more detailed overview on this topic, the readers are referred to~\cite{zhan2023multimodal}. Among all generative synthesis and editing methods, an important approach is Pix2PixHD~\cite{wang2018high}, which generates high-resolution synthetic images using conditional GANs from semantic label maps. Another influential work is SPADE~\cite{park2019semantic}, in which semantic image synthesis is achieved through spatially-adaptive normalization using a VAE. This method has been reported to generate better synthesis results compared to Pix2PixHD~\cite{wang2018high}. More recently, 
diffusion models have also been used for this task, with the latest examples including SDM~\cite{wang2022semantic} and CycleDiffusion~\cite{wu2023latent}. The former exploits the use of DDPM for semantic image synthesis, while CycleDiffusion investigates the stochastic diffusion probabilistic models in the latent space, and shows its effectiveness for various image editing tasks. Specific models have also been designed for image-to-image translation, including those based on GANs~\cite{lee2018diverse,sushko2020you}, autoregressive models~\cite{chen2020generative}, VAEs~\cite{esser2021taming,wu2022nuwa} and diffusion models~\cite{sun2023sddm,tumanyan2023plug}.

\subsection{Generative Video Frame Interpolation}

\begin{figure*}[ht]
\small
    \centering
    \begin{minipage}[c]{0.222\linewidth}
\centering
    \includegraphics[width=1\linewidth]{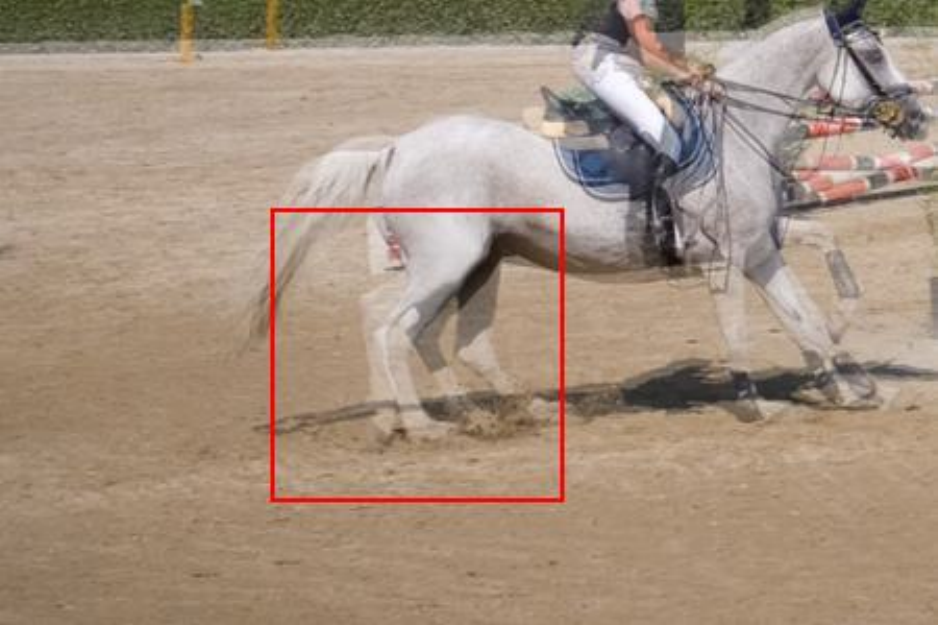}
     Overlayed Inputs
\end{minipage}
    \begin{minipage}[c]{0.15\linewidth}
\centering
    \includegraphics[width=1\linewidth]{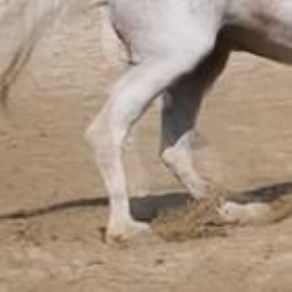}
    GT
\end{minipage}
    \begin{minipage}[c]{0.15\linewidth}
\centering
    \includegraphics[width=1\linewidth]{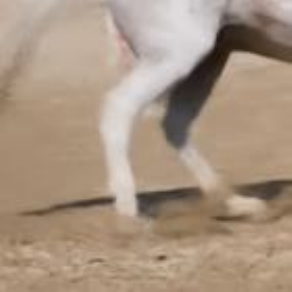}
    BMBC~\cite{park2020bmbc}
\end{minipage}
    \begin{minipage}[c]{0.15\linewidth}
\centering
    \includegraphics[width=1\linewidth]{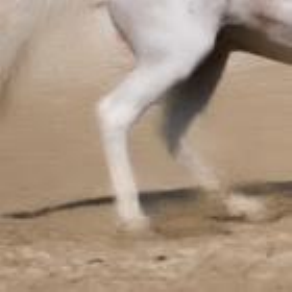}
      IFRNet~\cite{kong2022ifrnet}
\end{minipage}
    \begin{minipage}[c]{0.15\linewidth}
\centering
    \includegraphics[width=1\linewidth]{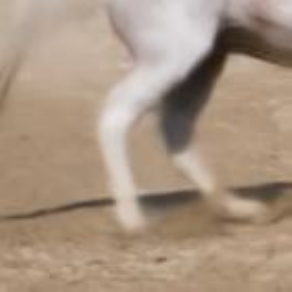}
    ST-MFNet~\cite{danier2022st}
\end{minipage}
    \begin{minipage}[c]{0.15\linewidth}
\centering
    \includegraphics[width=1\linewidth]{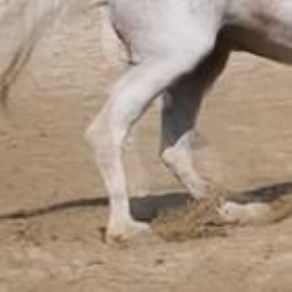}
    LDMVFI~\cite{danier2023ldmvfi}
\end{minipage}

    \begin{minipage}[c]{0.222\linewidth}
\centering
    \includegraphics[width=1\linewidth]{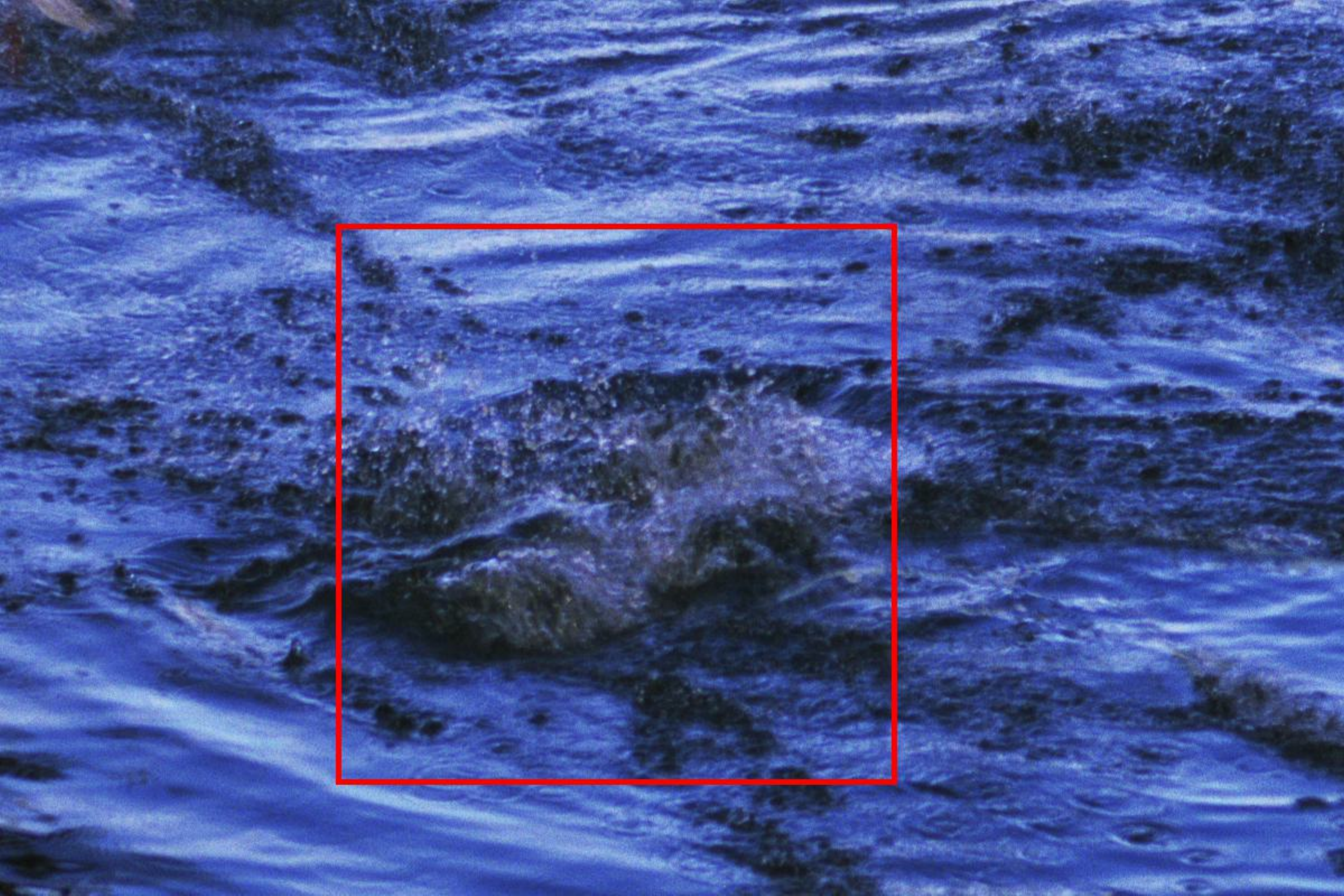}
    Overlayed Inputs
\end{minipage}
    \begin{minipage}[c]{0.15\linewidth}
\centering
    \includegraphics[width=1\linewidth]{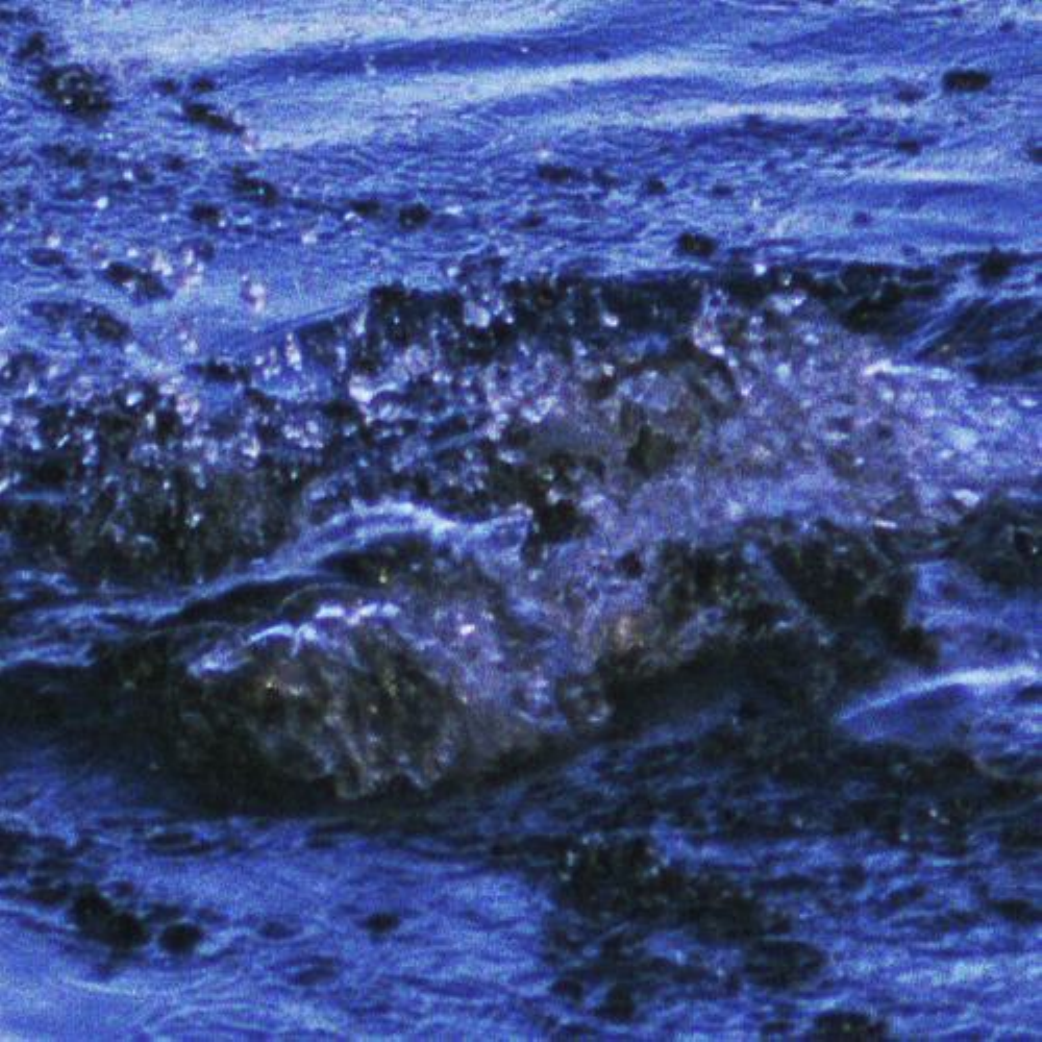}
    GT
\end{minipage}
    \begin{minipage}[c]{0.15\linewidth}
\centering
    \includegraphics[width=1\linewidth]{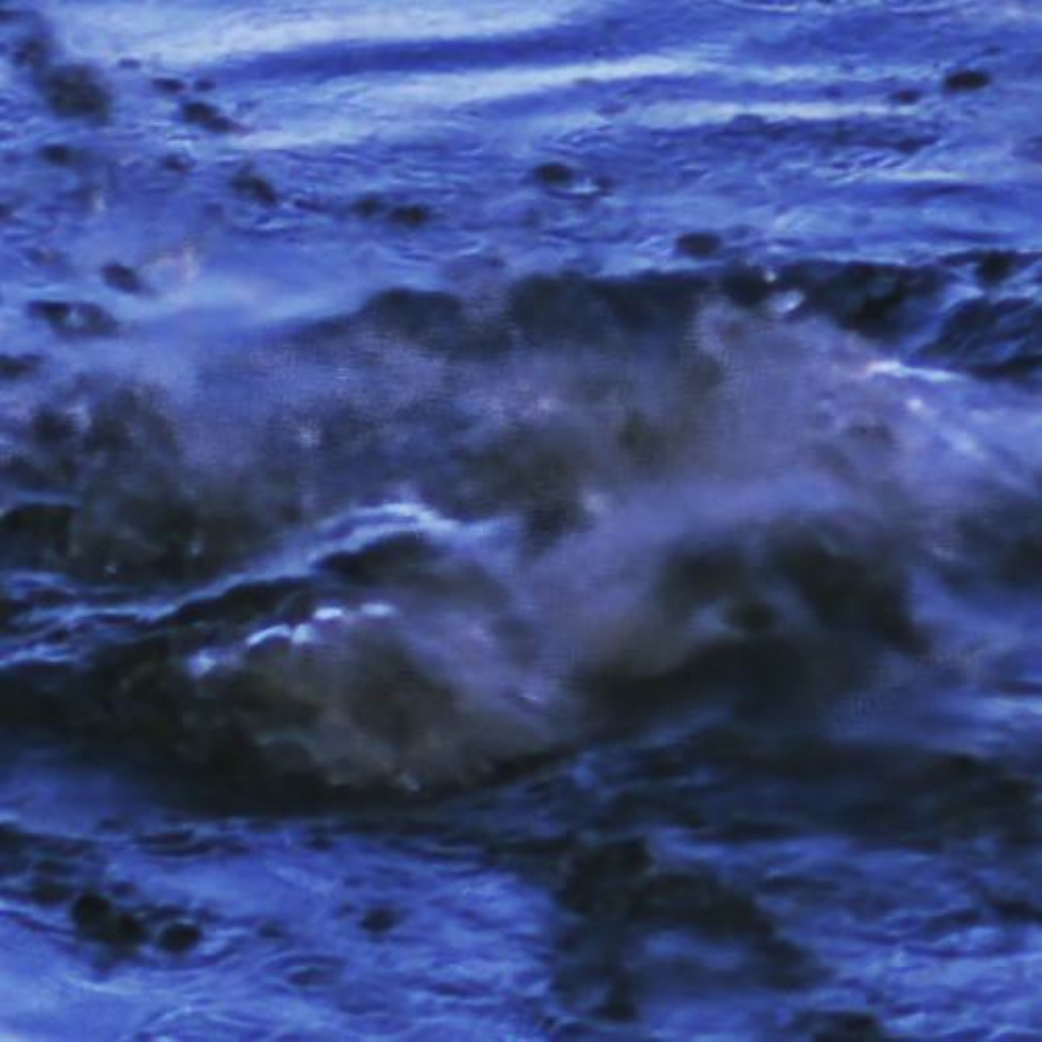}
     BMBC~\cite{park2020bmbc}
\end{minipage}
    \begin{minipage}[c]{0.15\linewidth}
\centering
    \includegraphics[width=1\linewidth]{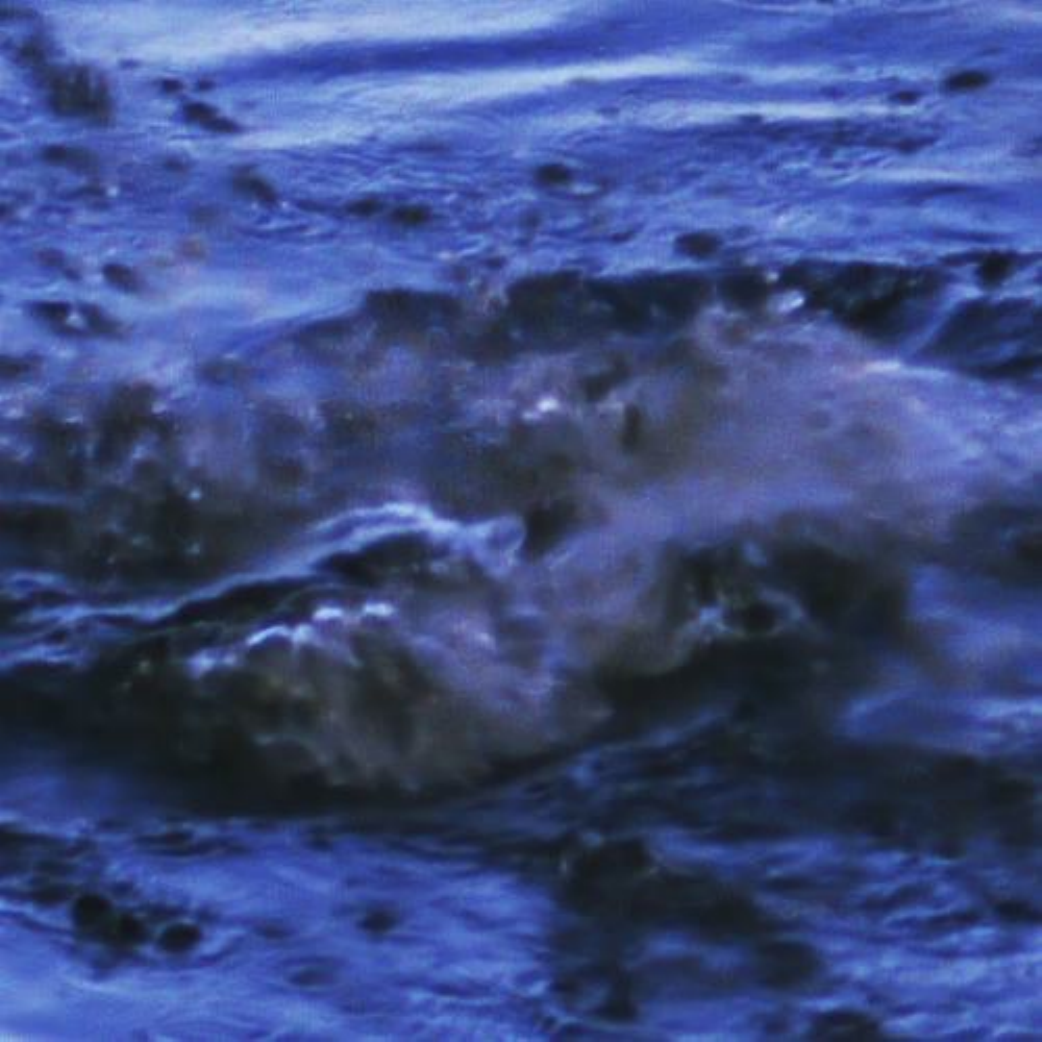}
    IFRNet~\cite{kong2022ifrnet}
\end{minipage}
    \begin{minipage}[c]{0.15\linewidth}
\centering
    \includegraphics[width=1\linewidth]{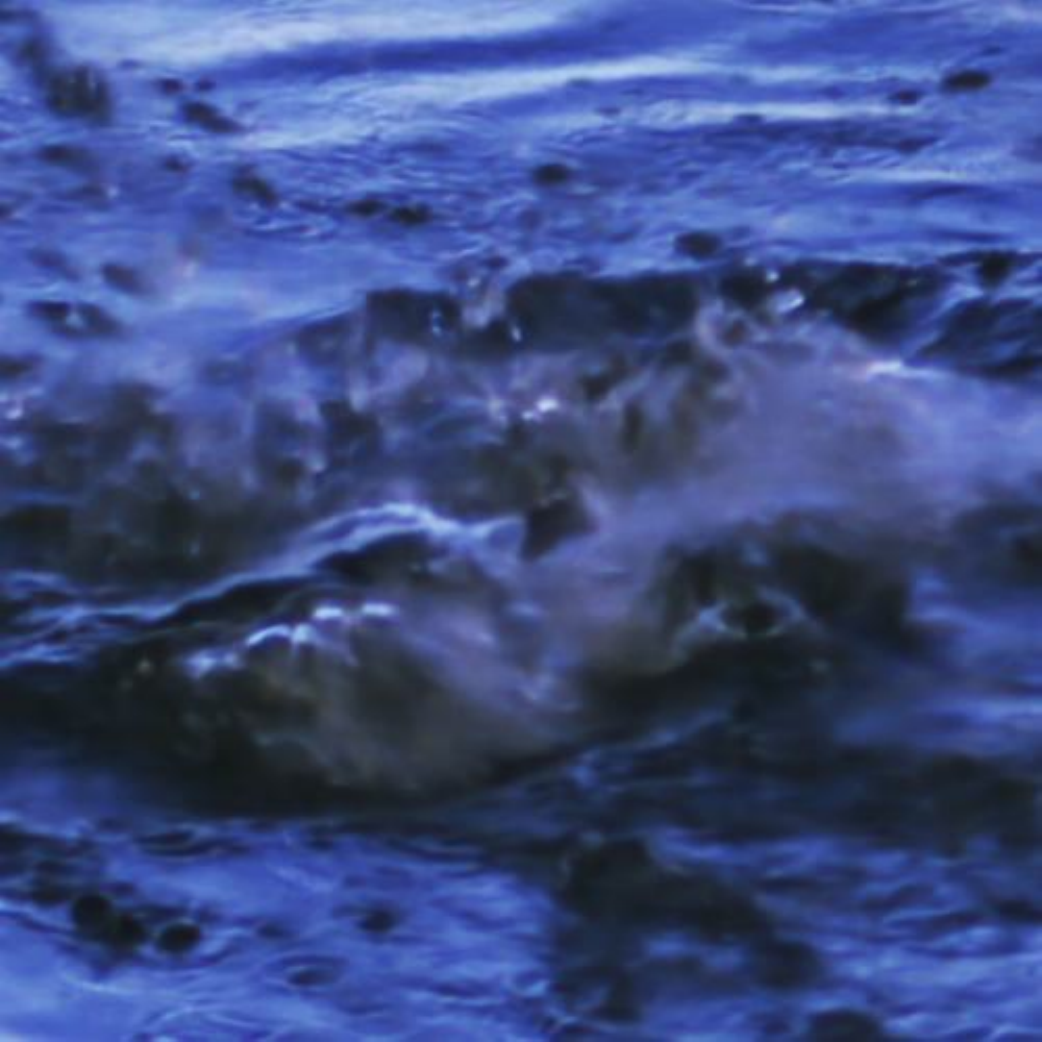}
    ST-MFNet~\cite{danier2022st}
\end{minipage}
    \begin{minipage}[c]{0.15\linewidth}
\centering
    \includegraphics[width=1\linewidth]{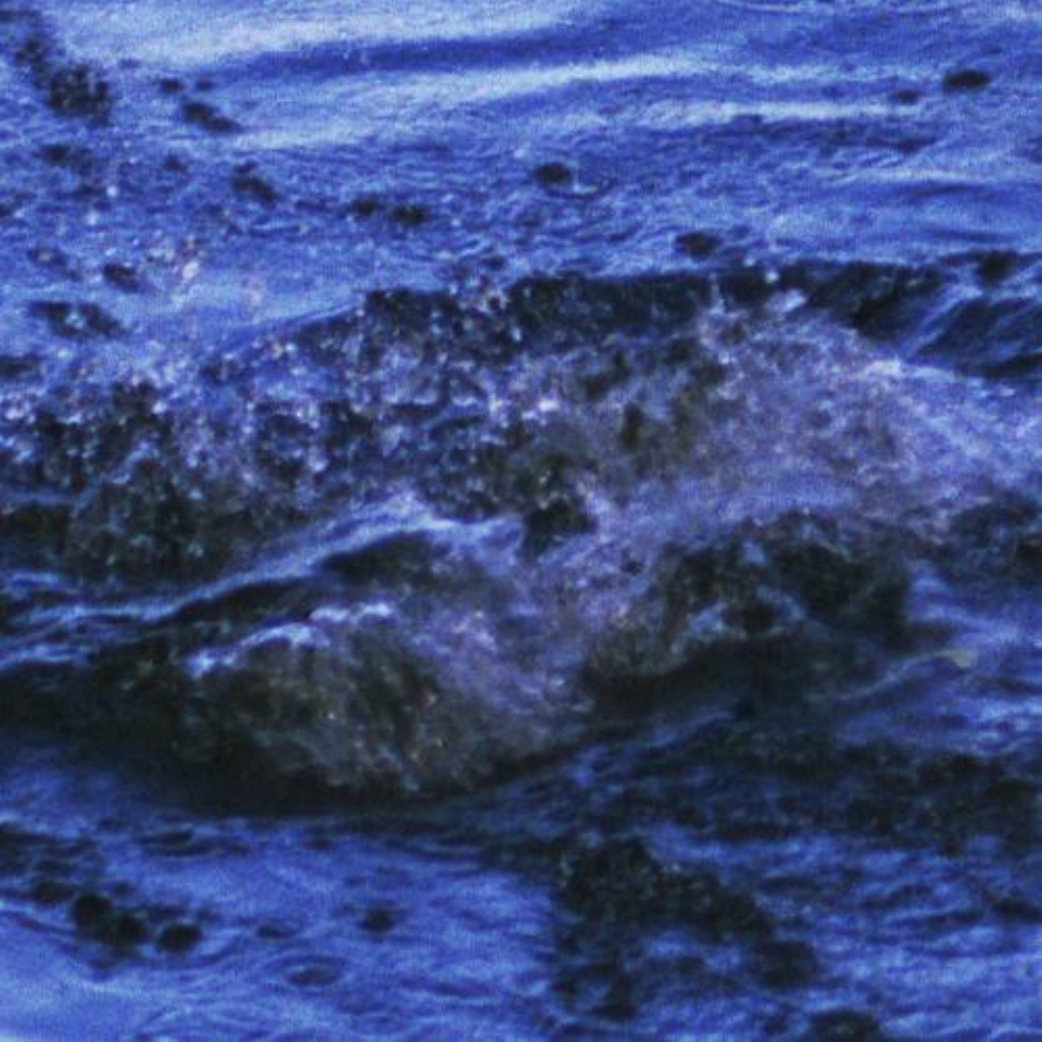}
    LDMVFI~\cite{danier2023ldmvfi}
\end{minipage}
    \caption{Visual comparison of the video frame interpolation results generated by SotA generative and non-generative models.}
    \label{fig:vfi}
\end{figure*}

Similar to super-resolution, which increases spatial resolution, video frame interpolation (VFI) is the technique for generating content with higher frame rates through synthetically creating intermediate frames between existing consecutive video frames. Non-generative leaning-based VFI methods are typically classified as kernel-based~\cite{niklaus2018context,sim2021xvfi} or flow-based~\cite{liu2017video,kong2022ifrnet}, based on different network structures and motion models, respectively. Recently, generative VFI methods have also been developed to obtain interpolated content with high-fidelity perceptual quality. GANs are widely adopted in this research field, typically with an adversarial network used to enhance the generator to produce results with better visual quality. The standard GAN architecture has been employed in~\cite{koren2017frame} for VFI; A multiscale GAN structure was developed in~\cite{van2017frame} to achieve improved visual quality and faster interpolation speed; Two GANs are concatenated in~\cite{wen2019generating} to learn spatial and temporal (motion) information separately. Moreover, Frame-GAN was developed based on Wasserstein GANs with gradient penalty~\cite{gulrajani2017improved} and a modified generator loss. Recently, the research of VFI has been boosted by the advances in diffusion models. One of the first diffusion-based VFI approaches is the Masked Conditional Video Diffusion (MCVD)~\cite{voleti2022mcvd}, which is based on a probabilistic conditional score-based denoising diffusion model. Its performance has been further outperformed by MV-Diffusion (Motion-aware Video Diffusion)~\cite{deng2023mv}, in which long-term and short-term motion trajectories have been learned using DDPMs with a motion trend attention model. More recently, latent diffusion models (LDM) have been adapted to VFI, and one of the resulting approaches, LDMVFI~\cite{danier2023ldmvfi}, achieves superior interpolation performance, in particular for video content with large motions and dynamic scenes. Fig. ~\ref{fig:vfi} shows the visual comparison results between LDMVFI and four other well-performing, non-generative VFI methods, IFRNet~\cite{kong2022ifrnet}, BMBC~\cite{park2020bmbc} and ST-MFNet~\cite{danier2022st}. It has been observed that LDMVFI can reconstruct sharp edges and textural details, which are similar to those in the ground truth content, while other methods tend to result in interpolation results with blurring and structural artifacts. This demonstrates the effectiveness of diffusion models when employed for VFI.

\section{Generative Quality Assessment}
\label{sec:QA}

In this section, we summarize the related work on generative model-based quality assessment and the quality metrics developed for assessing generative models.

\subsection{Generative Model-based Quality Assessment}

Visual quality assessment is one important research topic in visual signal processing. It is typically used for evaluating and comparing the performance of different processing methods. In the context of deep learning, quality metrics can be also employed in the training process as a loss function, to optimize the model generalization. In the current literature, generative models have also been used for image and video quality assessment. For example, a blind image quality metric has been proposed in ~\cite{yang2019blind} based on a GAN, which generates training samples to tackle a common issue with deep learning based quality assessment, the lack of reliable training content. A further work~\cite{ma2020active} focuses on using a GAN to predict the primary content of a distorted image, with internal generative mechanism inspired constraints. Moreover, alternative GAN architectures have been exploited by researchers in this research field, such as~\cite{zhu2021recycling} using  Wasserstein GANs to achieve opinion-unaware image quality assessment. Recently, diffusion models (DDPM in this case) have also been integrated in the image quality assessment framework for facial content~\cite{ziga2023diffiqa}, to generate perturbations and estimate their influence on perceptual quality.

\subsection{Quality assessment for Generative Models}

It is noted that generative models produce visual content which tends to have different characteristics compared to that generated by non-generative learning based and conventional processing approaches. This challenges the current common practices used for algorithm evaluation. It leads to another research area, quality assessment for generative models. To this end, researchers have developed various quality databases that contain content produced by generative models. This provides a ground-truth database for benchmarking existing quality metrics and developing new assessment methods. LCIQA~\cite{zhang2022no} is one of such works, which generates test content employing various learned video codecs based on commonly used CNNs and GANs. The resulting database was then used to evaluate several widely used full reference and no reference quality metrics. Another study~\cite{betzalel2022study}  investigates the performance of heuristic metrics such as the Inception Score (IS)~\cite{salimans2016improved} and the Fr'echet Inception Distance (FID)~\cite{heusel2017gans} for generative models in the context of image generation. The results show that although these metrics offer a relatively good correlation to several f-divergences, their ranking ability is limited when generative model performance is close. To further improve the quality prediction accuracy for generative models, enhancement methods have been proposed, including compound FID (CFID)~\cite{nunn2021compound}, GAN-IQA~\cite{ko2020quality} and DR-IQA~\cite{zheng2021learning}. A recent work in this domain proposed a lightweight generalizable framework to evaluate generative models~\cite{zhao2024lightweight}. The new metrics developed in this work demonstrate improved quality prediction performance compared to existing evaluation methods, such as FID~\cite{heusel2017gans}.

\section{Optimization of visual signal coding and processing with generative models}
\label{sec:Optimization}

Though generative models have shown promising results in visual signal coding and processing, their implementation requires care and optimization. First, due to the use of neural networks, the complexity is often too high to meet real-time requirements. To solve this problem, there have been several fast optimization techniques at the algorithmic and architectural levels. Second, issues such as the model robustness and variable bitrate also deserve investigation.

\subsection{Fast Optimization for Learned Image Compression}

In this section, we introduce various optimization techniques for learned image compression models that adopt the factorized prior and/or hyperprior.

\subsubsection{Algorithmic Optimization}

Network quantization plays an important role in algorithmic optimization. As shown in \cite{isscc}, compared with the 32-bit floating-point arithmetic, the 8-bit fixed-point arithmetic reduces the energy consumption for additions and multiplications by 30x and 19x, respectively. Therefore, network quantization is crucial for fast and low-complexity implementations. Different from other generative applications, learned codecs require entropy coding, which demands the bit-exact accuracy to ensure interoperability across platforms. As a result, network quantization is also essential to bit-exact computation. Several quantization methods have been proposed in the literature \cite{balle2018integer,sun2021learned,he2022post,koyuncu2022device,jeon2023integer,koyuncu2023quantized}.

As reported in \cite{gholami2022survey}, there are two major approaches to network quantization: (1) the quantization-aware training (QAT) and (2) the post training quantization (PTQ). For the QAT scheme, a pre-trained floating-point network is quantized and fine tuned in the presence of quantization. The fine tuning may be carried out with respect to the network weights and/or the additional parameters (e.g. quantization bit depth) for quantization. For the PTQ scheme, the pre-trained floating-point network is directly quantized. The key procedure is to use the calibration dataset to find the optimal clipping range and quantization schemes (e.g. linear or non-linear quantization with or without zero offset).

For learned image compression, \cite{jia2022fpx} exploits the Vitis AI Quantizer to generate an 8-bit quantized model. Both PTQ and QAT are supported in the Vitis AI Quantizer. After the quantization, the coding performance loss is tolerable in terms of the bpp overhead and PSNR loss. \cite{sun2021learned} proposes a channel-wise QAT scheme for the weight quantization. A heuristic fine-tuning scheme starting from the synthesis transform is developed. In the case of the 8-bit quantization, the coding loss is negligible compared with the 32-bit anchor. In order to further reduce the coding loss resulting from the activation quantization, a PTQ method based on the channel splitting is proposed in \cite{sun2022q}. Specifically, some channels with large magnitudes are equivalently split into multiple channels to reduce quantization errors while a few channels are pruned to maintain the overall network complexity. As compared with the previous work \cite{sun2021learned}, it achieves a BD-rate saving of up to 4.74\% . Another innovative work \cite{shi2023rate} proves that the well-used mean square error reduction is not an optimal criterion to decide the quantization parameters. Alternatively, they propose a rate-distortion optimized PTQ (RDO-PTQ), which uses the rate-distortion cost as the criterion for PTQ. Compared with \cite{sun2022q}, \cite{shi2023rate} performs better on MSE-optimized models.

In addition to network quantization, network pruning, another type of network compression, has also been applied to learned codecs. \cite{luo2022memory} aims at pruning the hyper path. Based on ResRep \cite{ding2021resrep}, a Lasso penalty is added in the loss function to adapt the number of pruned channels. Results show that at least 22.6\% of the network parameters are saved with a negligible coding loss. \cite{kim2020efficient} proposes an asymmetric framework composed of a heavy encoder and a lightweight decoder. In addition, the unstructured element-wise pruning and structured channel-wise pruning methods have been trialed. Interestingly, in the case of channel-wise pruning, removing the channels with larger $l_1$ norms is found to be more effective than removing the ones with smaller $l_1$ norms.

There have been some other low-complexity algorithms. To reduce the decoding complexity, \cite{guo2022evc} realizes a real-time framework by mask decay. They utilize knowledge distillation techniques to transform the parameters from large models to small models. By doing so, the coding performance is improved by more than 30\% for smaller models. In addition, the residual representation learning is proposed to implement a variable-rate encoder. \cite{zheng2021get} utilizes independent separable downsampling and upsampling components to reduce the network burden. Besides, similar to \cite{kim2020efficient}, an asymmetric architecture is proposed to boost the decoding speed. When implemented on Intel Core i7-9700K@3.6GHz, it reaches a decoding throughput of 37.5 FPS for small models.

In addition, some hardware-oriented algorithmic optimizations have also been proposed. \cite{yin2023bandwidth} assumes that activations (feature maps) dominate the data transfer between the on-chip and off-chip memory. To reduce the bandwidth for transferring the activations, they propose a differentiable pipeline to include the required bandwidth in the classical rate-distortion loss function for training. Compared with using only the rate-distortion cost as the loss function, this modified training objective incurs little loss in coding performance. \cite{yu2023lut} utilizes lookup tables to construct the hyper decoder. Compared with inferencing neural networks, both the model size and runtime are much reduced.

\subsubsection{Architectural Optimization}

This section further introduces some FPGA implementations. \cite{jia2022fpxnic} utilizes AMD/Xilinx Zynq UltraScale+ MPSoC ZCU104 fabricated with 16nm technology. The PL chip is XCZU7EV-2FFVC1156, which owns the resource of 504 kilo logic cell, 38 Mb memory and 1728 DSP. \cite{jia2022fpxnic} uses DPU as the hardware accelerator, and the working frequency is 350 MHz. It achieves 3.90 FPS for 720P, 1.68 FPS for 1080P and 0.42 FPS for 4K resolutions, respectively. Though the throughput is rather low, as one of very early FPGA-based learned codec frameworks, \cite{jia2022fpxnic} represents a starting point and offers useful insights into future research directions. It is a complete system, including video capturing, encoding, decoding and display. \tcb{The model of \cite{jia2022fpxnic} is mainly based on a block-wise coding framework\cite{9401164}.}

\begin{figure}[t]
	\centering
	\centerline{\includegraphics[width=9cm]{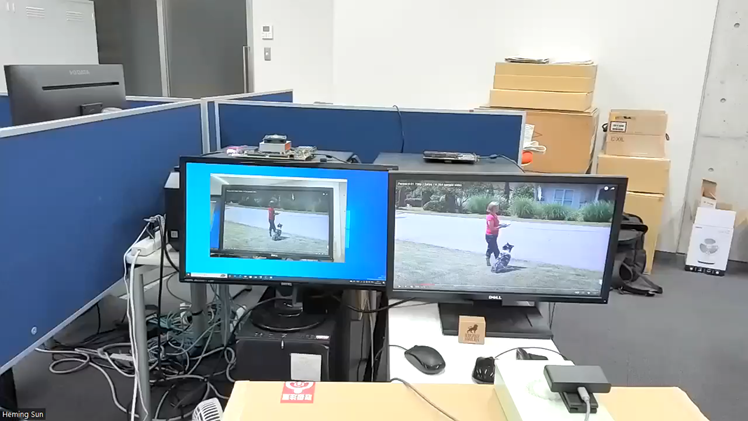}} 
	\caption{A demo system for learned image compression. Encoding is performed on an FPGA board KU115, while decoding is performed on an FPGA board VCU118. The links of demo videos are given in the footnote.}
	\label{fig_demo}
\end{figure}

\cite{shao2023high} also utilizes two UltraScale+ MPSoC evaluation boards ZCU102 and ZCU104 working at 200 MHz \tcb{to implement a factorized model \cite{balle2016end}}. The authors implement the hardware accelerator by Verilog HDL. When dealing with 256 $\times$ 256 images, it requires 15.87 ms and 14.51 ms for encoding and decoding, respectively. Note that to reduce the hardware complexity, \cite{shao2023high} also proposes a piece-wise linear approximation of the generalized divisible normalization (GDN) operation and its inverse operation. 
L eLe
\cite{sun2022f} proposes an FPGA architecture with a fine-grained pipeline \tcb{to implement the hyperprior model in \cite{cheng2019deep}}. Different from using DPU, which is a generic architecture, the proposed pipeline architecture is more flexible for the neural layers with various CTC ratios. When implemented with an AMD/Xilinx Virtex UltraScale FPGA VCU118 development board, it achieves 40.69 FPS and 35.77 FPS for encoding and decoding 720P videos, respectively. For 1080P videos, it achieves 19.15 FPS and 16.83 FPS for encoding and decoding, respectively. \cite{sun2022real} gives a CPU-FPGA system where entropy coding is performed at the CPU side and neural computing is performed at the FPGA side. A system-level pipeline is required to process the tasks on CPU and FPGA in a parallel manner. A demo system is given in Fig. \ref{fig_demo} where the encoding is performed on an FPGA acceleration board KU115 and the decoding is performed on VCU118. Some demo videos \tcb{can} be found here \footnote{\url{https://youtu.be/-unSbqsUS8Y}} \footnote{\url{https://youtu.be/Y4QO2h0LEDQ}}.

\color{black}{
For the above three FPGA codec systems, the power efficiency are around 29 GOPS/W, 47 GOPS/W and 21 GOPS/W for \cite{sun2022f}, \cite{jia2022fpxnic} and \cite{shao2023high}, respectively. If those GOPS/W satisfy the required performance of learned codecs, then those codecs can run smoothly on the devices. For example, assuming one watt power supply for the circuit performing the codec operations, we can only compute about 40 giga operations per second.

However, the above results are from FPGA implementations. As compared to FPGA implementations, ASIC implementations are expected to have much higher power efficiency. However, up to now, there has been no ASIC implementation for learned image compression. One potential reason is that traditional codecs have specific components such as intra/inter prediction and DCT, so that we are able to develop corresponding ASIC chips such as \cite{tsai20131062mpixels,huang2013249mpixel,zhou201614}. However, learned codecs are mainly based on neural networks, the computation of which can be executed efficiently on neural processing units (NPU). An edge device equipped with an NPU chip capable of delivering 20 TOPS/W may meet the computation requirements of some recent learned image codecs such as \cite{liu2023learned}, which consumes about hundreds of GOP for one Kodak image.

}
\color{black}

\subsection{Fast Optimization for Learned Video Compression}

For video compression, since the inter frame is taken into account, the complexity becomes even higher. As a result, fast optimization techniques become even more desirable.

\subsubsection{Algorithmic Optimization}

\cite{tian2023towards} presents a real-time design, reaching 720P@25FPS decoding on GeForce RTX 2080. To avoid the cross-platform interoperability issue, the coordinates of the transboundary quantization positions are included in the bitstream. Besides, several lightweight methods such as model pruning have been adopted to reduce the decoding complexity. As a result, decoding an I-frame takes 37.1 ms and decoding a P-frame takes 39.9 ms. With a Group of Pictures (GOP) of 12 frames, the average time per frame is 39.7 ms, which translates into 28.1 FPS. \cite{peng2023accelerating} proposes a novel model-agnostic pruning scheme based on gradient decay and layer-wise distillation. The effectiveness has been evaluated on various learned video codecs: FVC, DCVC and DCVC-HEM. As a result, 2$\times$ speed-up with less than 0.3 dB BD-PSNR loss is achieved.

\subsubsection{Architectural Optimization}

As an extension of \cite{jia2022fpxnic}, \cite{jia2022fpx} gives an FPGA implementation for learned video compression with P-frame. By using the same deployment methods and evaluation board, a P-frame framework is mapped onto FPGA. When tested on JVET Class B and Class C datasets, it achieves better coding performance than x264-veryfast.

\cite{zhang2023computationally} gives an ASIC design for learned video compression. \tcb{Based on a residual coding framework,} it features a CNN-Transformer neural network to enlarge the receptive field. Moreover, it adopts the Winograd algorithm to implement fast convolution and deconvolution. Notably, they develop a reconfigurable processing unit for the proposed fast algorithm. In addition, a dedicated data flow is presented to minimize the off-chip memory access. When synthesized by Synopsys Design Complier with TSMC's 28nm technology, it is able to operate at 400 MHz. With a throughput of 3525 GOPS,  the real-time decoding of a 1080P video at 25 FPS is made possible. Compared with CPU and GPU implementations, this design has significantly higher energy efficiency in terms of GOPS per Watt.

\subsubsection{System Optimization}

Mobilecodec \cite{le2022mobilecodec} is the first-ever real-time inter-frame learned video decoder. When tested on Snapdragon 8 chip, it achieves a decoding throughput of $>$30 FPS for 720P videos. Similar to traditional video compression, video frames are processed in the unit of GOP. For coding intra frames, a typical VAE-based model with the hyperprior is adopted. For coding inter frames, they follow the residual coding framework \cite{lu2019dvc}, which is composed of the motion network and the residual network. To reduce the complexity, a flow-agnostic motion compensation network with only convolutional operations is proposed.

To run with the fixed-point arithmetic, \cite{le2022mobilecodec} utilizes QAT to fine tune the 8-bit quantized network. Based on the dynamic range of the computation, the channel-wise quantization is adopted for both weights and activations. Regarding the computational complexity, the I-frame decoding costs 130.9 kMAC/Pixel, and the P-frame decoding consumes 257.1 kMAC/Pixel.

As an enhanced version of \cite{le2022mobilecodec}, \cite{van2024mobilenvc} realizes faster throughput and better coding performance. In detail, it reduces the decoding MAC by 10 $\times$ and saves 48\% BD-rate compared with \cite{le2022mobilecodec}. The same as \cite{le2022mobilecodec}, \cite{van2024mobilenvc} is also based on the residual coding framework. Different from \cite{le2022mobilecodec}, a block-based warping scheme is proposed for the P-frame coding. Regarding the network quantization, the symmetric channel-wise quantization is adopted for weights, whereas the asymmetric layer-wise quantization is adopted for activations. Here, the asymmetricity represents the use of a zero offset. \cite{van2024mobilenvc} also provides a system-level pipeline for the tasks on CPU, GPU, NPU and warping kernel. When tested on HEVC-B dataset, it is able to decode the videos at 38.9 FPS. Regarding the coding performance, there is still a gap between the 8-bit integer version and H.264 (FFmpeg).

\subsection{Fast Optimization based on Implicit Neural Representations}

\tcb{There are two types of Implicit Neural Representation (INR). One uses the pixel coordinates as inputs and the network learns to generate the color values for the pixel in question. Another simply overfits an autoencoder or a decoder for a given image/video.}

\tcb{For the first category, }starting from \cite{dupont2021coin}, there have been quite a few INR-based image/video compression methods \cite{strumpler2022implicit,kim2023c3,zhang2022implicit,ladune2023cool,van2021instance,dupont2022coin++,gomes2023video,damodaran2023rqat,guo2023compression,zhang2023enhanced}.

As one of the very first attempt at INR compression, \cite{dupont2021coin} overfits the image with a small MLP. After that, the weights of MLP are quantized and stored as the bitstream. Compared with the typical hyperprior frameworks, there is still room for improvement in terms of compression performance. However, the model size is only 14 kB, which is smaller than the hyperprior by several orders of magnitude.

\cite{strumpler2022implicit} tackles some issues of the design in \cite{dupont2021coin}. The first issue is the prolonged encoding time for overfitting the model. To accelerate the overfitting process, \cite{strumpler2022implicit} uses the idea of meta learning to determine the initial weights. The second issue is the inferior coding performance. To solve this problem, the post-quantization optimization and entropy coding have been proposed for the INR-based compression framework. As a result, it outperforms \cite{dupont2021coin} significantly in terms of the coding performance. The convergence speed is also much faster than \cite{dupont2021coin}.

As a very recent work, \cite{kim2023c3} proposes a combined manner of latent and INR-based compression. INR works for two networks: the entropy network and synthesis network. At the decoding side, the entropy network and the synthesis network are reconstructed based on the decoded weights. The means and scales are then generated to formulate the prior distribution. Through the entropy decoding, the latent is recovered. Finally, the latent will be sent to the synthesis network to generate the decoded image. In addition to image compression, \cite{kim2023c3} also extends the framework to video compression. As a result, \cite{kim2023c3} only consumes 3 kMAC/Pixel and 5 kMAC/Pixel for image and video, respectively. Furthermore, its coding performance is quite attractive. For image coding, it is comparable with VVC. For video, it is comparable with \cite{mentzer2022vct}.

\tcb{For the second category, the key concept is to use overfitting to reduce the amortization gap \cite{cremer2018inference}. There have been several works \cite{van2021overfitting,balcilar2022reducing,jourairi2022improving,9675360,balcilar2022reducing1}. \cite{jourairi2022improving} improves the reconstruction quality by overfitting the bias in the decoder. \cite{van2021overfitting} fine tunes not only the encoder and latent, but also the entire model. \cite{balcilar2022reducing} makes the factorized prior and hyperprior models more suitable to the test instance. \cite{9675360} studies how to overfit some important parameters in order to reduce the overhead. Different from the above works for images, \cite{balcilar2022reducing1} fine tunes the model for video.}

\subsection{Other Optimization Schemes for Practicality Considerations}

The above three subsections are mainly for fast and low-complexity implementations. In addition to the speed, there are several optimization techniques for practicality considerations. We mainly introduce the adversarial attack and variable bitrate in this subsection.

Adversarial attack is one way to mislead the results of neural networks. For learned codecs, since the neural network structure is usually disclosed, the attacker can easily fetch the network parameters and generate adversarial inputs. The target of the adversarial attack can be the reconstructed quality (e.g. PSNR, MS-SSIM) or bitrate. There have been several works targeting at the attack and defense of learned codecs \cite{liu2023manipulation,song2024training,zhu2024attack,chen2023towards}. \cite{liu2023manipulation} is a very early attempt at the white-box and black-box attack on learned image compression. \cite{song2024training} proposes a training-free defense framework with a random input transform. The method does not influence the rate-distortion result for the clean image. \cite{zhu2024attack} gives a comprehensive study on various attack methods, attacking targets, neural network structures and bitrates. The attack transferability to VVC was also studied in \cite{zhu2024attack}. \cite{chen2023towards} also tries various settings for the attack. Besides, several efficient defense methods such as pre-processing and adversarial training are proposed.

Variable bitrate is another important issue. Different from the traditional codec which utilizes quantization parameters (QPs) to control the bitrate, learned codecs usually incorporates a hyperparmeter $\lambda$ to adjust the trade-off between rate and distortion. Each $\lambda$ corresponds to a specific model, which increases the storage cost of network models. To solve this problem, there have been several works \cite{choi2019variable,cui2021asymmetric,song2021variable}. \cite{choi2019variable} adopts a conditional autoencoder. $\lambda$ and quantization bin size are used for the rate control. \cite{cui2021asymmetric} adopts multiple $\lambda$ in the training phase, so that the resulting model is able to interpolate between the pre-trained $\lambda$ to achieve variable-rate coding. \cite{song2021variable} utilizes a quality map to generate prior condition features, and then insert these features into the encoder and decoder to realize variable-rate coding.

\section{Conclusion}
\label{sec:Conclu}

This paper offers a concise review of generative models, along with a comprehensive survey of visual signal coding with generative models, focusing specifically on algorithms for image and video coding. Additionally, it addresses the recent advancements in international standardization efforts for visual signal coding with generative models. These efforts are extremely important for media industry. 
This paper also discusses the research works of applying generative models to various image and video processing tasks, such as restoration, synthesis, editing, and interpolation, along with visual signal quality assessment using generative models and quality assessment for generative models. Finally, this paper discusses the latest advancements in optimization research on visual signal coding and processing with generative models. 
The field of visual signal coding and processing with generative models is vast and rapidly evolving, making it difficult to undertake a comprehensive overview that includes all relevant works. Inevitably, some important research or emerging trends may have been overlooked in this paper. We hope that this survey will provide valuable insights and encourage further exploration and innovation among researchers in this field.

\small
\bibliographystyle{IEEEtran}
\bibliography{IEEEabrv,ref}

\end{document}